\documentclass{aa}  

\usepackage{txfonts}

\usepackage{helvet}
\usepackage{graphicx}
\usepackage{array}
\usepackage{float}
\usepackage{color}
\usepackage[utf8]{inputenc}  
\usepackage{amsmath}  
\usepackage{comment}
\usepackage{footnote}
\usepackage{longtable}
\usepackage{ccaption}
\usepackage{url}
\usepackage[hyphenbreaks]{breakurl}
\usepackage[figuresright]{rotating}
\usepackage{xurl}
\usepackage{soul, xcolor}
\bibpunct{(}{)}{;}{a}{,}{,}

\newcommand{\degree}{\ensuremath{^\circ}}

\usepackage{hyperref}
\hypersetup{colorlinks=true,breaklinks=true,citecolor=blue}
\usepackage{graphicx}	
\usepackage{amsmath}	
\setstcolor{red}
\begin{document}

   \title{Optical polarimetry study of Lambda-Orionis star-forming region}


   \author{Sharma Neha
          \inst{1},
          Archana Soam\inst{2}
          \and
          G. Maheswar\inst{2}
          }

   \institute{Aryabhatta Research Institute of Observational Sciences (ARIES), Nainital, 263002, India\\
              \email{pathakneha.sharma@gmail.com \& neha.astro18@gmail.com}
         \and
             Indian Institute of Astrophysics (IIA), Sarjapur Road, Koramangala, Bangalore 560034, India
             }

   \date{accepted May 20, 2024}

 \abstract{
 We present an optical polarimetric study of a nearby star-forming region, Lambda-Orionis, to map plane-of-the-sky magnetic field geometry to understand the magnetized evolution of the HII region and associated small molecular clouds. We made multi-wavelength polarization observations of 34 bright stars distributed across the region. R-band polarization measurements focused on small molecular clouds BRC 17 and BRC 18 located at the periphery of the HII region are also presented. The magnetic field lines exhibit a large-scale ordered orientation consistent with the Planck sub-mm polarization measurements. The magnetic field lines in both the BRCs are found to be roughly in north-south directions; however, a larger dispersion is noticed in the orientation for BRC 17 compared to BRC 18. Using structure-function analysis, the strength of the plane-of-the-sky component of the magnetic field is estimated as $\sim$28 $\mu$G for BRC 17 and $\sim$40 $\mu$G for BRC 18. The average dust grain size and the mean value of the total-to-selective extinction ratio (R$_{V}$) in the HII region are found to be $\sim$0.51 $\pm$ 0.05 $\mu$m and $\sim$2.9 $\pm$ 0.3, respectively. The distance of the whole HII region is estimated as $\sim$392 $\pm$ 8 pc by combining astrometry information from GAIA EDR3 for YSOs associated with BRCs and confirmed members of central cluster Collinder 69.}
 

   \maketitle
%


\section{Introduction} \label{sec:intro}
HII regions are among the most fascinating astronomical objects at optical wavelengths. Several analytical and numerical studies have been conducted to comprehend their formation and expansion (see \citealt{2011MNRAS.414.1747A}, \citealt{2019MNRAS.487.2200Z} and reference therein) as well as to explain the filamentary structures, globules, pillars, clumps, etc. found within and around these regions (e.g., \citealt{1991ApJS...77...59S}, \citealt{2009MNRAS.396..964C}). These features could be due to the instabilities in the ionization front or the density inhomogeneities in the molecular cloud inside which the HII region is expanding (see \citealt{1996ApJ...469..171G}, \citealt{2010MNRAS.403..714M}). Small, isolated, and dense molecular clouds, called bright-rimmed clouds (BRCs), are located on the periphery of the evolved HII regions. When HII regions interact with their surrounding molecular clouds, they form the BRCs. The ultraviolet radiation from the nearby OB stars photoionizes the outer surface layers of a BRC, forming a layer of hot ionized gas known as the ionized boundary layer (IBL). The recombination processes in the IBL result in the formation of optically bright rims. The photoionization-induced shocks are driven into the molecular clouds, forming dense cores that are then triggered to collapse to form new stars \citep{2011EAS....51...45E}. This phenomenon of triggered star formation by the compression of gas due to shock or ionization front is known as radiatively driven implosion (RDI; \citealt{1989ApJ...346..735B}; \citealt{2006MNRAS.369..143M}) scenario.

Magnetic field plays a very crucial role in the molecular clouds by regulating the star formation processes \citep[see][]{McKeeOstriker2007}. The expanding HII region and photo-dissociation region (PDR) tend to remove pre-existing small-scale disordered magnetic field structure and produce a large-scale ordered magnetic field in the neutral shell, with approximately parallel orientation to the ionization front \citep{2011MNRAS.414.1747A}. \cite{2013ApJ...774..128S} found magnetic field lines are preferentially oriented parallel to most of the density structures, and the relative orientation changes from parallel to perpendicular if the density increases over a critical density. The strength of the magnetic field decides whether the field lines will align themselves along the direction of ionization radiation or remain unchanged (\citealt{2009MNRAS.398..157H}, \citealt{2011MNRAS.412.2079M}).

The role of the magnetic field has been investigated in several BRCs using polarimetry. Optical polarimetric studies have shown that the magnetic field could shape the BRCs along with the ionizing radiation from the hot star(s) located in the vicinity of the BRCs \citep{2017MNRAS.465..559S, 2018MNRAS.476.4782S}. The ALMA dust polarization study of NGC 6334 by \citet{Cortes_2021} reveals complex morphology with a significant degree of turbulence, which could be due to the strong interactions between the magnetic field and the turbulent gas in the BRC. Recent sub-millimeter observations, such as POL-2 on the James Clerk Maxwell Telescope (JCMT), have revealed complex magnetic fields in regions where stars are yet to form \citep{Liu_2019, Pattle_2021}.

The Orion molecular cloud complex is a nearby giant molecular cloud complex whose head part is known as ``The Lambda Orionis Star Forming Region (LOSFR)." It is one of the unique locations to study the young stellar population and their environment \citep{2008hsf1.book..757M}. The LOSFR is an HII region \citep[Sh2-264]{1959ApJS....4..257S} illuminated by an O8III type star called $\lambda$-Ori or Meissa at a distance of 450 $\pm$ 50 pc from the Sun \citep{1999AJ....118.2409D}. Another B0 type main-sequence star is in binary with Meissa at an angular distance of 4.4$^{\prime\prime}$. \citet{1989A&A...218..231Z} detected a dense molecular gas and dust ring of a diameter of 9$\degree$ centered around $\lambda$-Ori star using Infrared Astronomical Satellite (IRAS). This ring coincides with the shell of neutral hydrogen previously discovered by \cite{1957AJ.....62..148W}. A compact open cluster, Collinder 69, is located at the center of the LOSFR, and the $\lambda$-Ori star is the brightest member of this cluster. At the periphery of the $\lambda$-Ori ring, many dark clouds are located, e.g., Barnard 30 (B30), Barnard 35 (B35), Barnard 36 (B36), and Barnard 224 (B224) cataloged by \cite{Barnard1919}. Of these dark clouds, B30 and B35 are also termed as BRC 17 or SFO 17 and BRC 18 or SFO 18, respectively, cataloged by \cite{1991ApJS...77...59S}. The BRC 17 is located at the northwest edge separated from $\lambda$-Ori by $\sim$2$\degree$, and the elongated BRC 18 cloud is projecting inward from the eastern edge at an angular distance of $\sim$2.5$\degree$ from the $\lambda$-Ori.

The $\lambda$-Ori regions have been extensively studied in various aspects, e.g., the initial discovery, the distance of the clouds, an H$\alpha$ survey, analysis of IRAS data, and photometric and spectroscopic studies. \cite{1982ApJ...261..135D} performed an H$\alpha$ survey and identified emission line sources predominantly distributed near the open cluster Collinder 69 and two dark clouds, BRC 17 and BRC 18. \cite{1999AJ....118.2409D, 2001AJ....121.2124D, 2002AJ....123..387D} studied the whole LOSFR using Lithium survey and deep VRI photometry, concluding that young stars are spatially distributed around $\lambda$-Ori and in front of the BRC 17 and BRC 18. However, some young stars are found outside these high stellar density clouds, proposing that the star formation was centered in an elongated cloud extending from BRC 18 through $\lambda$-Ori to the BRC 17 cloud \citep{1999AJ....118.2409D}. The LOSFR comprises recently formed stars from 0.2 M$_{\odot}$ to 24 M$_{\odot}$ and still, the star formation continues in the dark clouds, BRC 17 and BRC 18 \citep{2008hsf1.book..757M}.

The magnetic field morphology in $\lambda$-Ori region will provide useful information about the expansion and evolution of LOSFR and the globules associated with this region. In this paper, we have mapped the global magnetic field geometry in multi-wavelength and local magnetic field geometry in BRC 17 and BRC 18 in R-band. The structure of this paper is as follows: in Sect. 2, we briefly describe the optical polarimetric observations and data reduction for both. Our results and analyses are presented in Sect. 3 followed by a discussion in Sect. 4. We present a summary and highlight our main findings in Sect. 5.

\begin{figure*}
\centering
\resizebox{8.0cm}{11cm}{\includegraphics{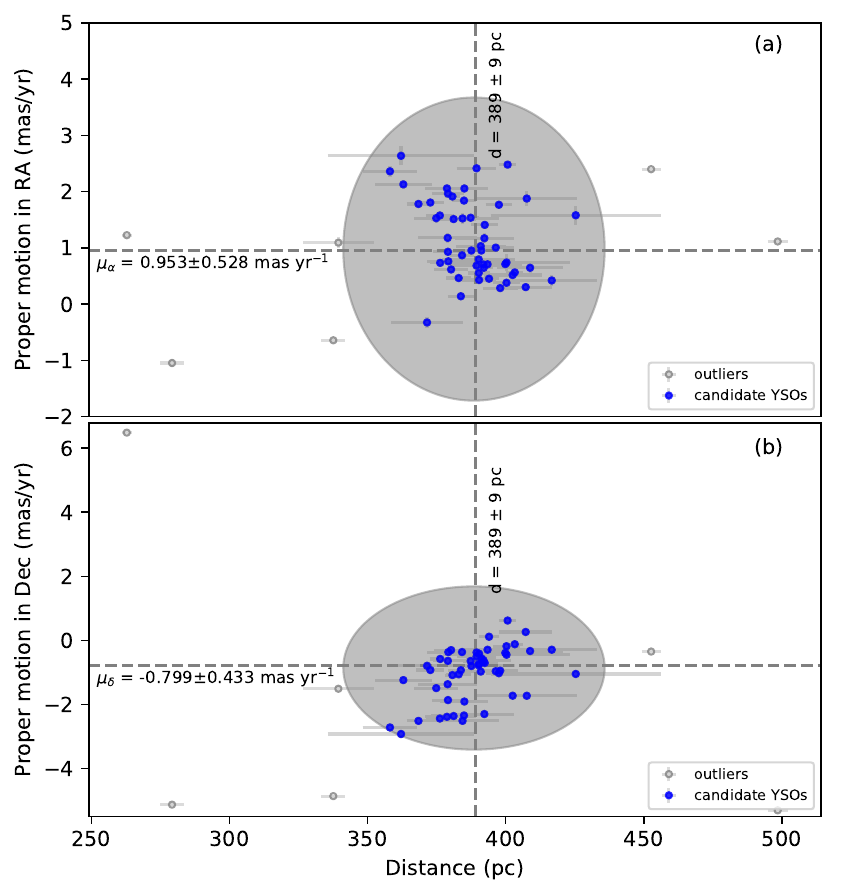}}
\resizebox{8.0cm}{11cm}{\includegraphics{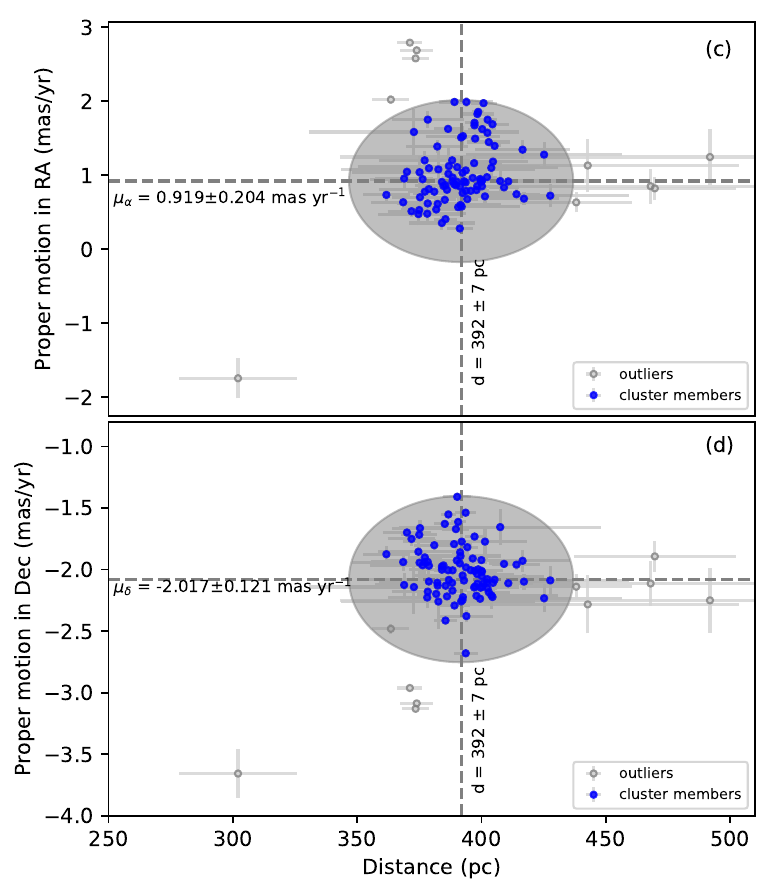}}
\caption{\small Proper motion values of the known candidate YSOs associated with BRC 17 (left panel) and the known probable cluster members associated with Collinder 69 (right panel) are plotted as a function of their distances obtained from Gaia EDR3. (a) \& (c): $\mu_{\alpha}$ versus d plot of the sources. The candidate YSOs/cluster members lying within 5 × MAD boundary of $\mu_{\alpha}$ versus d plot are plotted using filled blue circles, while the outliers are plotted using open gray circles. The dashed lines indicate the median values of $\mu_{\alpha}$ and d of the candidate YSOs/cluster members. (b) \& (d): $\mu_{\delta}$ versus d plot of the sources. The gray shaded ellipses show the 5 x MAD boundary range in distance and proper motions within which sources are used to estimate a distance. The candidate YSOs/cluster members lying within 5 × MAD boundary of $\mu_{\delta}$ versus d plot are plotted using filled blue circles, while the outliers are plotted using open gray circles. The dashed lines indicate the median values of d and $\mu_{\delta}$ of the candidate YSOs/cluster members.} \label{fig:BRC17_Co69_gaia}
\end{figure*}

\section{Observation and Data Reduction}

\begin{table}
\centering
\caption{Log of observations.}\label{tab:obslog}
\begin{tabular}{c c}\hline
Cloud ID & Date of observations (year, month, date)  \\ \hline
BRC 17        & 2014 December 17, 18, 20  	 \\
              & 2015 January 17, 19  		 \\
              & 2015 October 11  			 \\
              & 2015 November 15  			 \\ \hline
BRC 18        & 2014 October 19, 21, 22, 23  \\
              & 2014 November 17, 18, 19     \\
              & 2014 December 17, 20, 22   	 \\ \hline
$\lambda$-Ori & 2015 November 16, 17         \\
 Region  	  & 2015 December 14, 15         \\ 
			  &2016 January 8, 9, 10, 11, 12 \\ 
			  & 2016 February 8, 9           \\ 
			  & 2016 March    8				 \\ \hline
\end{tabular}
\end{table}
 

We performed the optical linear polarization observations of the LOSFR containing BRC 17 and BRC 18 using the ARIES IMaging POLarimeter (AIMPOL) mounted at the 104-cm Sampurnanand telescope, which is situated at Aryabhatta Research Institute of Observational Sciences (ARIES), Nainital, India. 
We used standard Johnson VRI filters having $\lambda_{\mathrm{Veff}}$ = 0.550 $\mu$m, $\lambda_{\mathrm{Reff}}$ = 0.670 $\mu$m, and $\lambda_{\mathrm{Ieff}}$ = 0.805 $\mu$m for the polarimetric observations. A detailed description of AIMPOL and the polarization measurement methods are given in \citet{1998A&AS..128..369R, 2004BASI...32..159R}. The steps of data reduction methods are described in previous papers \citep[e.g.,][]{2011MNRAS.411.1418E, 2012MNRAS.419.2587E, 2013A&A...556A..65E, 2016A&A...588A..45N, 2018MNRAS.476.4442N, 2013MNRAS.432.1502S, 2015A&A...573A..34S, 2017MNRAS.465..559S, 2018MNRAS.476.4782S}. The observations were performed between 2014 and 2016. The details of the observations are given in the Table \ref{tab:obslog}. We have observed the stars that are brighter than 15 magnitudes in the V-band towards each region in LOSFR.

Unpolarized standard stars were observed during every run to check for any possible instrumental polarization. The instrumental polarization is found to be less than 0.1$\%$. We corrected the observed degree of polarization for the instrumental polarization. The reference direction of the polarizer was determined by observing polarized standard stars from \cite{1992AJ....104.1563S}. We observed these unpolarized and polarized standards using the standard Johnson VRI filters. \cite{1992AJ....104.1563S} used standard Johnson V filter and Kron-Cousins R and I filters for the observations of the standard stars. We used the offset between the standard polarization position angle values estimated from the observations and those taken from \cite{1992AJ....104.1563S} to correct the zero point offset. The zero point offset values for polarization position angles range from 6$\degree$ to 9$\degree$ for all the observations.

\section{Results and analyses}

\subsection{Distance to the LOSFR} \label{distance}
We have used GAIA EDR3 data to estimate the distance of LOSFR \citep{2021A&A...649A...1G}. We separately estimated the distances towards BRC 17, BRC 18, and central cluster Collinder 69. The detailed distance estimation for BRC 18 is presented in \citet{2022MNRAS.510.2644S}. In the case of BRC 17, we have compiled a list of YSOs from literature using the criterion given in \citet{2022MNRAS.510.2644S}. We selected candidate YSOs towards BRC 17 within a circular region of 0.5$\degree$ radius around the IRAS sources (\citealt{2001AJ....121.2124D}, \citealt{2012PASJ...64...96H}, \citealt{2015AJ....150..100K}, \citealt{2018AJ....156...84K}, \citealt{2018A&A...620A.172Z}, \citealt{2019IJAA....9..154H}). In the case of Collinder 69, we have selected probable cluster members (\citealt{2012A&A...547A..80B}). After compiling all the sources, we searched for each source within a 1" search radius and obtained distance, proper motions in right ascension and declination from \cite{Bailer_Jones_2021} and Gaia EDR3 archive (\citealt{2021A&A...649A...1G}). We selected the sources having the ratio, x/$\Delta$x $\geq$ 3, where x represents the distance (d), proper motion in right ascension ($\mu_{\alpha}$) and declination ($\mu_{\delta}$) values and $\Delta$x shows their respective errors. We also applied the condition for renormalized unit weight errors (RUWE; \citealt{LL:LL-124, 2021A&A...649A...2L}) for all the sources. We selected the sources having RUWE $\leq$ 1.4 confirming their authentic astrometric measurements (\citealt{LL:LL-124}). The ionizing source $\lambda-$Ori for LOSFR is located in the central Collinder 69 cluster. The distance of $\lambda-$Ori is estimated at 386 $\pm$ 60 pc using GAIA EDR3 parallax measurements, but the RUWE is found to be 4.823, which is much higher than the cut-off value of 1.4 \citep{2021A&A...649A...1G}. Hence, we have used all the probable cluster members to estimate the distance of Collinder 69 and assign the same distance to $\lambda-$Ori. 

Fig. \ref{fig:BRC17_Co69_gaia} represents $\mu_{\alpha}$ and $\mu_{\delta}$ of the candidate YSOs associated with BRC 17 and the known cluster members associated with Collinder 69 as a function of their distances (d). In this study, we have estimated median absolute deviation (MAD) to measure the statistical dispersion in the data sets since the standard deviations are generally affected by extreme values. We identified the sources lying within 5 × MAD concerning the median values of the distance and the proper motions for BRC 17 and Collinder 69. Note that the distance uncertainties of individual sources have not been considered while calculating the median distance. After applying this condition, we excluded outliers and estimated the median and MAD values only for the sources within the above condition. The median with their respective MAD values for $\mu_{\alpha}$, $\mu_{\delta}$ and d are written in Table \ref{tab:gaia_dis} for BRC 17, BRC 18, and Collinder 69. The distances towards all three different regions of LOSFR are found to be consistent within error limits. Hence, we have taken the mean values of all three distance values to assign a single distance for the LOSFR. The mean distance is 392 $\pm$ 8 pc for the LOSFR.

\begin{table}
\centering
\caption{The median with their respective MAD values for $\mu_{\alpha}$, $\mu_{\delta}$ and d.}\label{tab:gaia_dis}
\begin{tabular}{c c c c}\hline
ID      & $\mu_{\alpha}$ & $\mu_{\delta}$ & d  \\ 
        & (mas yr$^{-1}$)& (mas yr$^{-1}$)& (pc)\\  \hline
BRC 17      & 0.953 $\pm$ 0.528 & -0.799 $\pm$ 0.433 & 389 $\pm$ 9\\ 
BRC 18$^{\dagger}$  & 2.189 $\pm$ 0.382 & -2.482 $\pm$ 0.114 & 394 $\pm$ 7\\ 
Collinder 69& 0.919 $\pm$ 0.204 & -2.017 $\pm$ 0.121 & 392 $\pm$ 7\\ \hline
\end{tabular}
$^{\dagger}$ The values for BRC 18 has been taken from \cite{2022MNRAS.510.2644S}.
\end{table}

\cite{1977MNRAS.181..657M} estimated the distance of the LOSFR using high-quality broadband photometry of the 11 OB stars located at the center of the star-forming region. They estimated the distance of this region to be $\sim$400 pc. The distance of the five stars in the central area of LOSFR is $\sim$380 $\pm$ 30 pc \citep{1997A&A...323L..49P}. \cite{2001AJ....121.2124D} used Str\"{o}mgren photometry with a larger sample of OB stars and estimated the distance of the region $\sim$450 $\pm$ 50 pc which is slightly larger than the previously estimated distances. Recently, \cite{2018AJ....156...84K} estimated the distances for BRC 17 as 397 $\pm$ 4 pc, BRC 18 as 396 $\pm$ 4 pc and Collinder 69 as 404 $\pm$ 4 pc using GAIA DR2 parallax measurements. Our estimated distances for BRC 17 and BRC 18 are consistent within error limits, but there is a small difference in the distance of Collinder 69. That could be because we have used the recent GAIA EDR3 distance measurements having very low uncertainties.

\subsection{Polarimetry}

\begin{figure}
\centering
\resizebox{8.5cm}{12cm}{\includegraphics{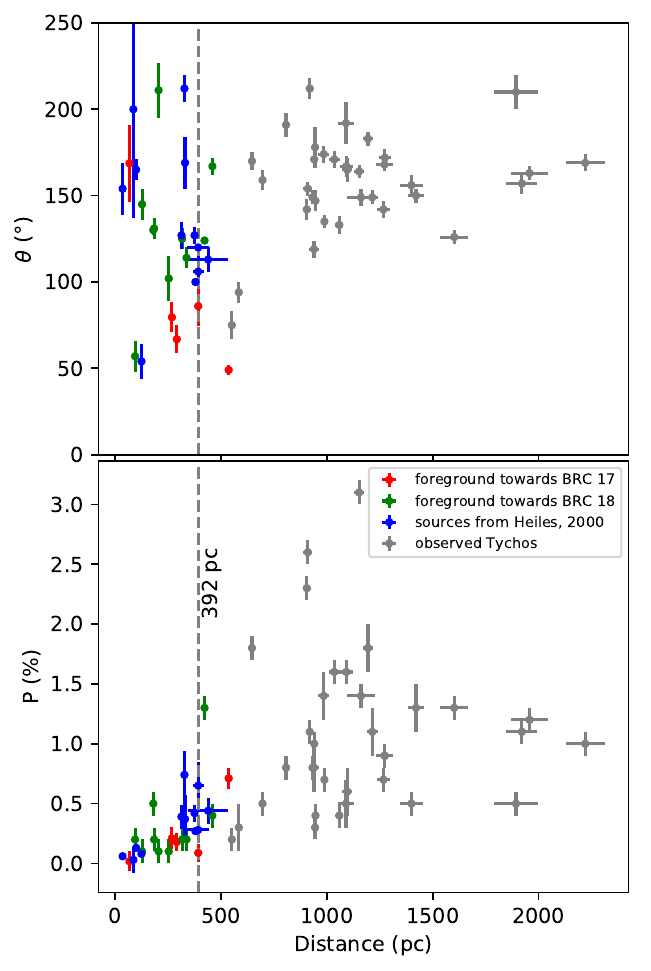}}
\caption{\small Upper panel: Polarization angle versus distance of the foreground stars found within $1\degree$ diameter around BRC 17, BRC 18 and the Heiles stars found within $9\degree$ diameter around $\lambda$-Ori star. The distances are calculated from the Hipparcos parallax measurements taken from \citet{2007A&A...474..653V} and adopted from \citet{Bailer_Jones_2021}. The gray circles show the bright Tycho stars lying behind the cloud. The distances for the Tycho stars are taken from \citet{Bailer_Jones_2021}. Lower panel: Degree of polarization versus distance for the same stars. The gray dashed line corresponds to the 392 pc distance in both panels.}\label{fig:P_theta_for}
\end{figure}

\subsubsection{Foreground polarization subtraction}\label{ch4:for}
The distance of the LOSFR is found to be 392 $\pm$ 8 pc as described in section \ref{distance}. Hence, the polarization measurements for the stars towards LOSFR could be affected by the line of sight foreground interstellar component. In this regard, to remove the foreground contribution, we searched BRC 17 and BRC 18 within a circular region of 1$\degree$ diameter for stars having parallax measurements made by Hipparcos satellite in \citet{2007A&A...474..653V}, since the GAIA data-sets were not available at the time of these observations. The stars that are found to be the emission line sources in binary or multiple systems or peculiar sources in SIMBAD are excluded. We have chosen the stars having parallax measurements better than 2$\sigma$. The polarization observations of these stars were carried out using AIMPOL. We have also chosen the stars from Heiles catalog \citep{2000AJ....119..923H} within 4.5$\degree$ radius of $\lambda$-Ori region. The polarization results for these stars have been presented in Table \ref{tab:table_fg_lambdaori} according to their increasing distance from the Sun. Note that the observed R-band polarization values are presented for BRC 17 and BRC 18, while the values from \citet{2000AJ....119..923H} catalog are the average values in multiple bands as taken directly from the catalog. We have also adopted distances for all the sources from \citet{Bailer_Jones_2021} to compare with previous known distances. We have plotted the degree of polarization and polarization angles with distance for all the foreground stars in Fig. \ref{fig:P_theta_for}. In the same figure, we have also included the polarization measurements for the bright Tycho stars \citep[][discussed in section \ref{global}]{2000A&A...355L..27H} that are lying behind the cloud. We can see from the figure that the values of polarization angle are random for the stars lying before 392 pc while aligned in a particular direction after 392 pc. Similarly, the value of the degree of polarization also shows an increase of around 392 pc, which is consistent with our measured distance. Hence, we calculated the mean values of P and $\theta$ for the foreground stars having a distance less than 392 pc according to a new distance estimation method \citep{Bailer_Jones_2021} and having polarization values greater than 0.1$\%$ which is equal to the instrumental polarization of AIMPOL. We have marked those objects with asterisks. The observed foreground R-band polarization values are converted to V-band and I-band using Serkowski law \citep{1975ApJ...196..261S}. The mean values of the P and $\theta$ are 0.3 $\pm$ 0.1 $\%$ and 106 $\pm$ 9$\degree$ for all the V, R and I filters. Using these mean polarization values, we calculated the Stokes parameters, Q$\mathrm{_{fg}}$ and U$\mathrm{_{fg}}$ for selected foreground stars. We also computed the Stokes parameters for our observed stars Q$_*$ and U$_*$. Then we calculated the foreground-corrected Stokes parameters for our observed stars Q$_{c}$ and U$_{c}$ using Q$_{c}$ = Q$_*$ $-$ Q$\mathrm{_{fg}}$ and U$_{c}$ = U$_*$ $-$ U$\mathrm{_{fg}}$ following the same procedure described in \cite{2016A&A...588A..45N}. However, the mean of the difference between observed polarization angles $\theta\mathrm{_{observed}}$ and foreground corrected polarization angles $\theta\mathrm{_{c}}$ is $~$4 degrees with a standard deviation of 4 degrees, suggesting a smaller contribution due to foreground. We have presented the polarization position angle values in the equatorial coordinate system.

\begin{table}
\centering
\caption{Polarization values of observed foreground stars and from Heiles catalog.}\label{tab:table_fg_lambdaori}
\scriptsize
\begin{tabular}{llcrrrr}\hline
\textbf{Id}& \textbf{Star Name}	& \textbf{V} & ${P\pm~\epsilon_P}$ & ${\theta \pm \epsilon_{\theta}}$  &\textbf{D}$^{\dagger}$ & \textbf{D}$^{\ddagger}$ \\ 
  &				&\textbf{(mag)}	& \textbf{(\%)} &\textbf{({$\degree$})}	&\textbf{(pc)} & \textbf{(pc)} \\\hline
\multicolumn{7}{c}{\bf{Observed foreground stars in the direction of BRC 17 {\bf in R-band}}} \\  
1  & HD 36860 & 8.0&  0.02 $\pm$ 0.1  & 169 $\pm$ 22 &  63$\pm$4  & 67$\pm$1 \\
2*  & HD 35968 & 8.0&  1.21 $\pm$ 0.1  & 80 $\pm$ 9 &  199$\pm$35 & 268$\pm$2 \\
3*  & HD 36104 & 7.0&  0.18 $\pm$ 0.1  & 67 $\pm$ 8 &  307$\pm$55 & 291$\pm$7 \\ 
4  & HD 36262 & 7.6&  0.09 $\pm$ 0.1  & 86 $\pm$ 12 &  333$\pm$72 & 392$\pm$7 \\ 
5  & HD 36838 & 8.2&  0.71 $\pm$ 0.1  & 49 $\pm$ 3 &  458$\pm$229 & 536$\pm$7 \\ \hline
\multicolumn{7}{c}{\bf{Observed foreground stars in the direction of BRC 18 {\bf in R-band}}}\\
 1*&  HD 38527 & 5.8&   0.17 $\pm$ 0.1 &  57 $\pm$ 9&   91$\pm$4 & 94$\pm$1 \\
 2*&  HD 39007 & 5.8&   0.15 $\pm$ 0.1 &  144 $\pm$9&  110$\pm$8 & 127$\pm$1\\ 
 3*&  HD 37408 & 8.6& 0.52 $\pm$ 0.1 &  130 $\pm$ 3&  178$\pm$45 & 180$\pm$1 \\
 4*&  HD 37355 & 6.7& 0.24 $\pm$ 0.1 &  131 $\pm$ 6&  208$\pm$29 & 184$\pm$1 \\
 5&  HD 246922& 8.9& 0.43 $\pm$ 0.1 &  167 $\pm$ 5&  274$\pm$120 & 459$\pm$4 \\
 6&  HD 37926 & 8.0& 0.04 $\pm$ 0.1 &  102 $\pm$ 13& 291$\pm$87 & 253$\pm$2 \\
 7&  HD 38095 & 8.5& 0.08 $\pm$ 0.1 &   31 $\pm$ 16& 335$\pm$159 & 205$\pm$1 \\
 8*&  HD 38096 & 8.0& 0.21 $\pm$ 0.1 &  125 $\pm$ 6&  351$\pm$92 & 316$\pm$3 \\
 9*&  HD 37522 & 7.2& 0.21 $\pm$ 0.1 &  114 $\pm$ 6&  385$\pm$135 & 337$\pm$3 \\
10&  HD 39229 & 8.7& 1.31 $\pm$ 0.1 &  124 $\pm$ 1&  387$\pm$125 & 422$\pm$4 \\ \hline
\multicolumn{7}{c}{\bf{Stars from \citet{2000AJ....119..923H}}}\\ 
1 &  HD 37160 & 5.8& 0.06 $\pm$ 0.03& 154 $\pm$ 15& 36$\pm$1 & 35$\pm$1 \\
2 &  HD 38899 &	4.9& 0.03 $\pm$ 0.12& 20  $\pm$ 63& 76$\pm$2 & 86$\pm$1 \\
3* &  HD 35468 &	1.6& 0.30 $\pm$ 0.21& 23  $\pm$ 20& 77$\pm$3 &  - \\
4* &  HD 36267 & 4.1& 0.13 $\pm$ 0.03& 165 $\pm$ 6 & 93$\pm$6 & 99$\pm$2 \\
5* &  HD 36653 &	5.6& 0.74 $\pm$ 0.20& 32  $\pm$ 8& 137$\pm$7 & 327$\pm$11 \\
6 &  HD 38710 &	5.3& 0.09 $\pm$ 0.03& 54  $\pm$ 10& 165$\pm$25 & 124$\pm$2 \\
7* &  HD 38672 & 6.7& 0.37 $\pm$ 0.20& 169 $\pm$ 15& 276$\pm$38 & 330$\pm$4 \\
8 &  HD 36822 &	4.4& 0.28 $\pm$ 0.00& 120 $\pm$ 0& 333$\pm$28 & 392$\pm$53 \\
9 &  HD 36861 &	3.5& 0.44 $\pm$ 0.11& 113 $\pm$ 7& 337$\pm$62 & 439$\pm$95 \\ 
10*&  HD 37232 &	6.1& 0.27 $\pm$ 0.00& 100 $\pm$	0& 345$\pm$52 & 379$\pm$9 \\
11*&  HD 36471 &	8.7& 0.42 $\pm$ 0.07& 127 $\pm$	5& 391$\pm$160 & 375$\pm$9 \\
12&  HD 34989 &	5.8& 0.65 $\pm$ 0.20& 106 $\pm$	9& 437$\pm$69 & 392$\pm$26 \\
13*&  HD 36824 &	6.7& 0.39 $\pm$ 0.10& 127 $\pm$	8&1176$\pm$817 & 313$\pm$5 \\ \hline
\end{tabular}

$^{\dagger}$ Distances are estimated using the Hipparcos parallax measurements taken from \citet{2007A&A...474..653V}.\\
$^{\ddagger}$ Distances are adopted from \citet{Bailer_Jones_2021}.\\
* Objects used to calculate foreground polarization. See details in the text.
\end{table}

\subsubsection{Global magnetic field geometry of LOSFR} \label{global}
\begin{figure}
\centering
\resizebox{0.45\textwidth}{7.1cm}{\includegraphics{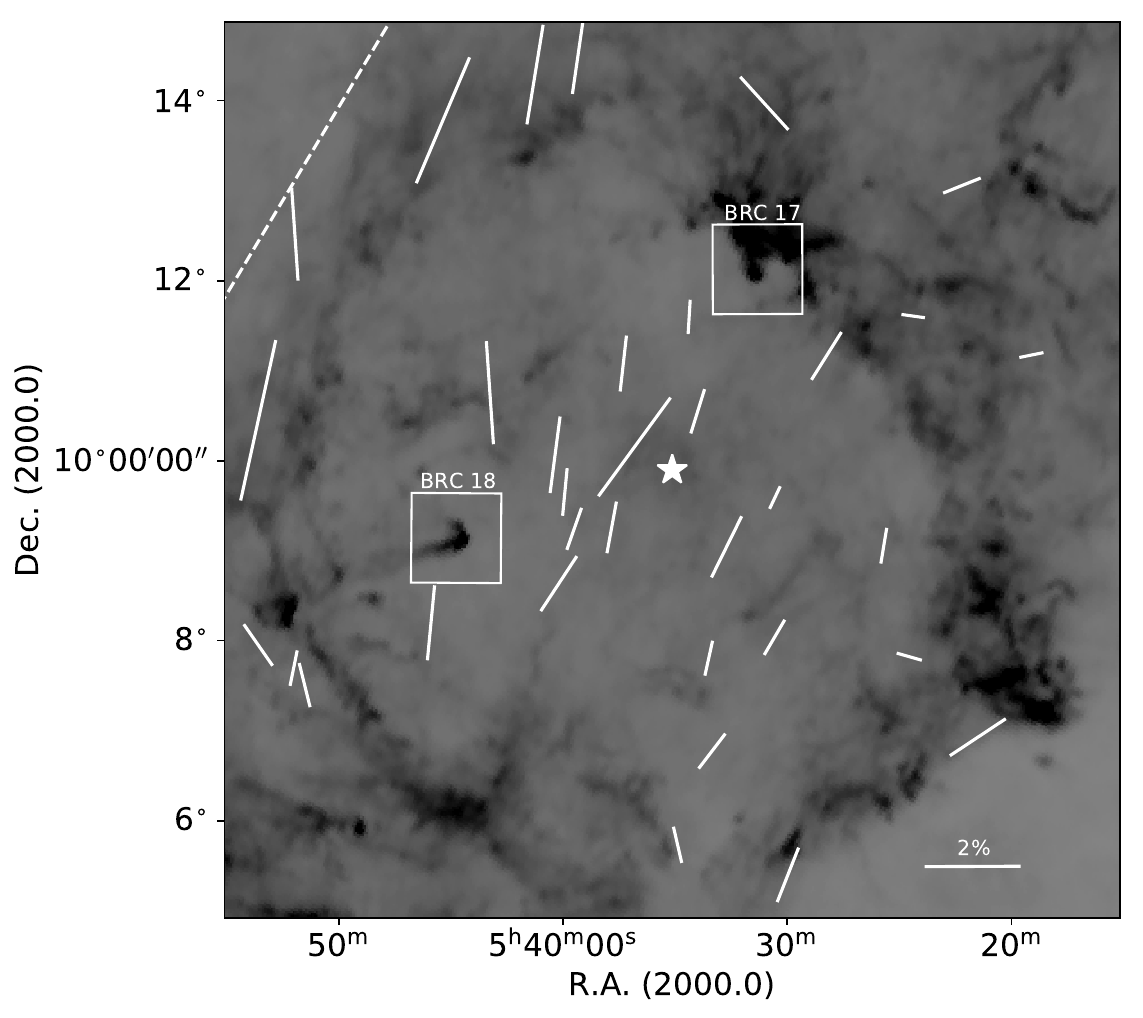}}
\resizebox{0.45\textwidth}{7.1cm}{\includegraphics{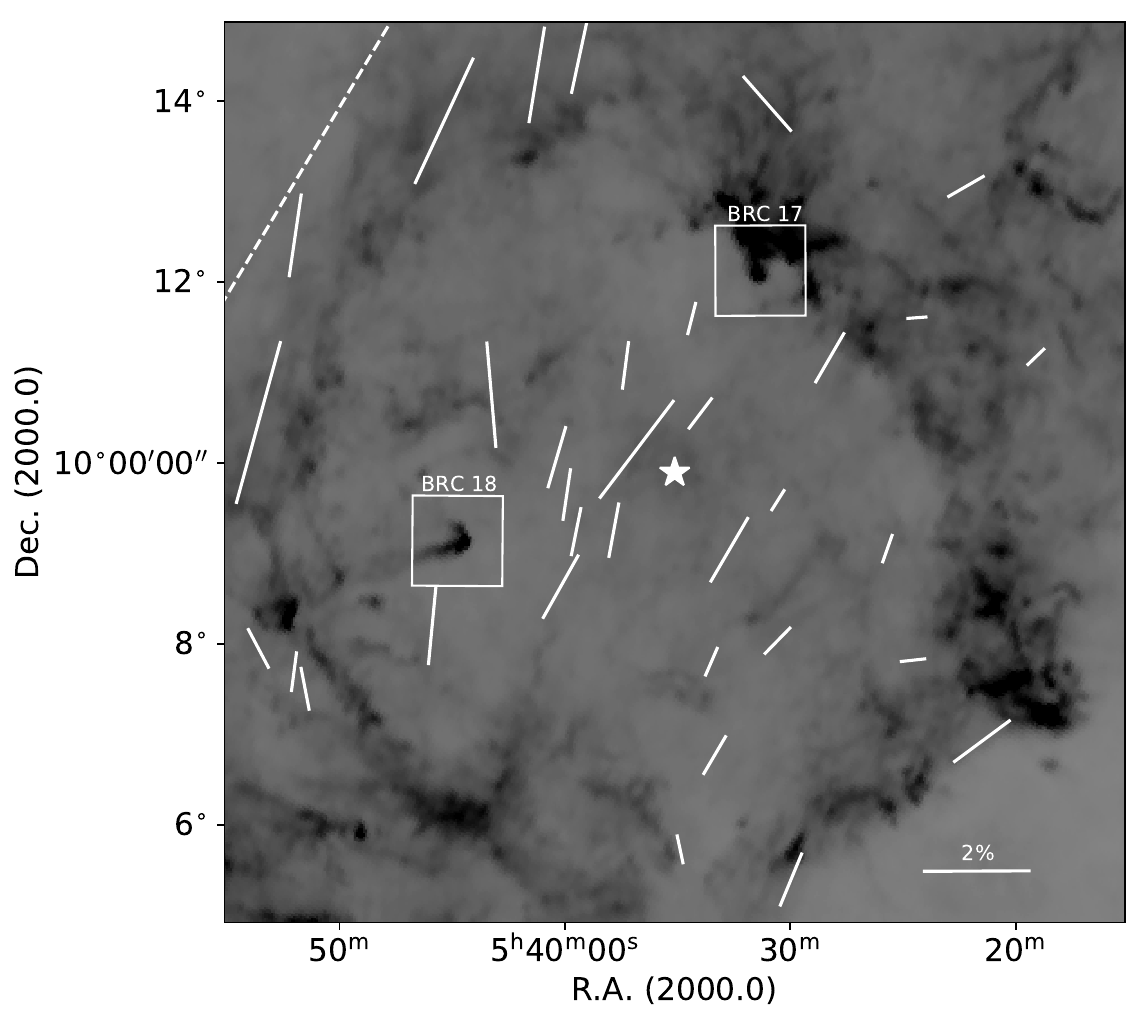}}
\resizebox{0.45\textwidth}{7.1cm}{\includegraphics{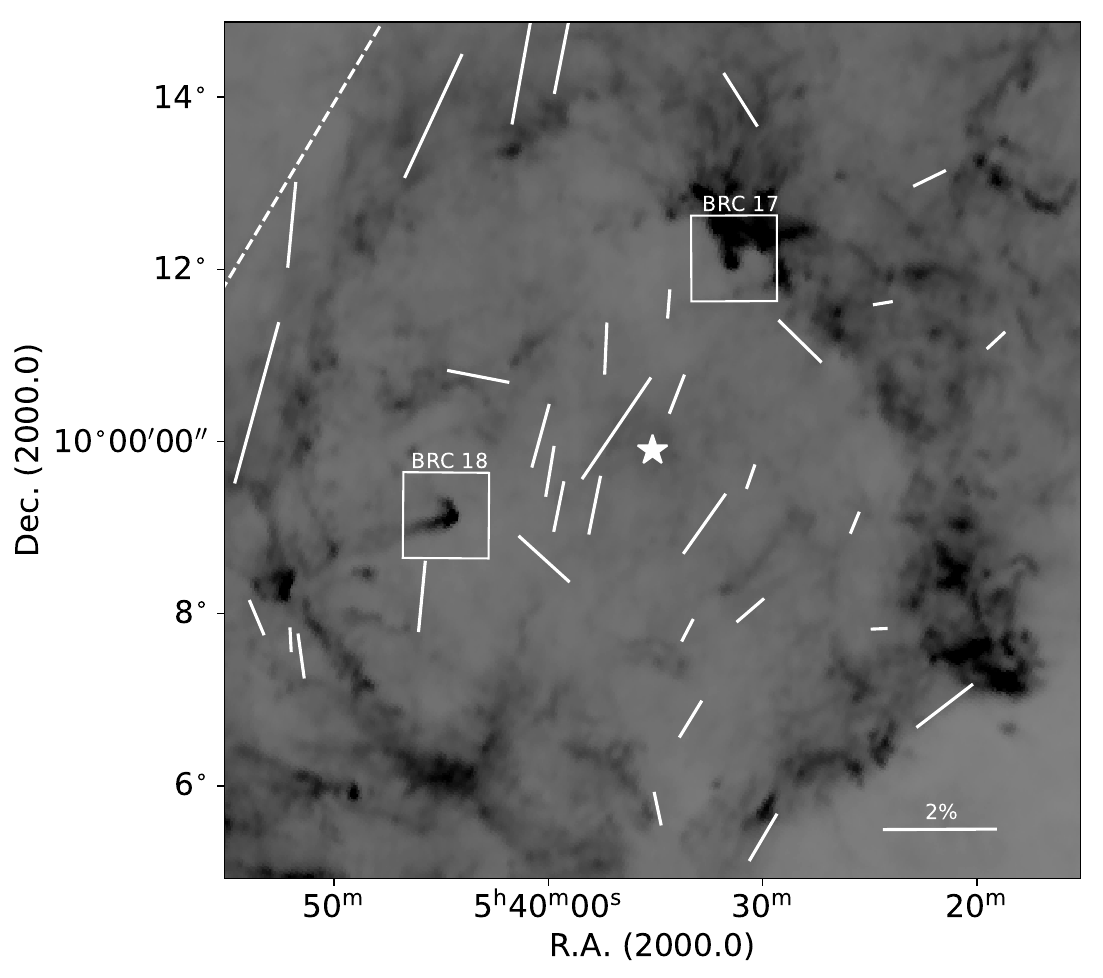}}
\caption{Polarization vectors of 34 bright Tycho stars plotted on the Planck 857 GHz image of LOSFR in V-band (Upper panel), R-band (Middle panel) and I-band (Lower panel). The length of the vectors corresponds to the degree of polarization values, while the orientation represents the polarization angle measured from the north toward the east. A vector with 2\% polarization has been shown for reference. The white dashed line shows the position of the galactic plane. The white star shows the position of $\lambda$-Ori star.}\label{fig:V_band}
\end{figure}

We have systematically made the polarization study to map the global magnetic field in the LOSFR by observing the bright Tycho stars located behind this region. We selected the bright Tycho stars from the Tycho-2 catalog \citep{2000A&A...355L..27H} within a 4.5$\degree$ radius around $\lambda$-Ori star. Then we have taken the J, H, and K$_{\mathrm{s}}$ magnitudes of these stars from 2MASS point source catalog \citep{Cutrietal2003}. We plotted their J, H, and K$_{\mathrm{s}}$ magnitudes in a color-color diagram and selected the stars with J$-$K$_{\mathrm{s}} >$ 1, showing the stars are reddened or behind the cloud. Among them, we made multi-band polarimetric observations of 34 bright Tycho stars distributed in the whole HII region. We have also obtained the distances for all these 34 bright Tycho stars from \cite{Bailer_Jones_2021} and found that all have larger distances than the LOSFR's distance (400 pc adopted from literature). This suggests that our method of selecting the reddened stars is reliable. We have plotted the polarization vectors on the Planck 857 GHz image in Fig. \ref{fig:V_band} for V, R, and I filters from top to bottom, respectively. The direction of polarization vectors shows the global magnetic field geometry in LOSFR, which seems to follow the large-scale structure seen in all the images of Fig. \ref{fig:V_band}. However, the number of stars in our sample is low; therefore, observations of more stars would be helpful. The white star shows the position of $\lambda$-Ori star in all the images, and the white dashed line represents the position of the galactic plane.
The magnetic field lines appear almost parallel to the Galactic plane (b = -11.99$\degree$) in the northeast and central regions of LOSFR, while towards the southwest region, the field lines seem random. The polarization results for all the Tycho stars towards LOSFR in VRI filters are given in Table \ref{tab:Tychos}. The mean values of the degree of polarization and polarization angle for these bright Tycho stars are 1.3 $\pm$ 0.2$\%$ \& 169 $\pm$ 6$\degree$ for V-band, 1.1 $\pm$ 0.1$\%$ \& 163 $\pm$ 6$\degree$ for R-band and 1.0 $\pm$ 0.1$\%$ and 170 $\pm$ 6$\degree$ for I-band, respectively. 

\subsection{Dust grain size distribution}

\begin{figure*}
\centering
\includegraphics[width=18cm]{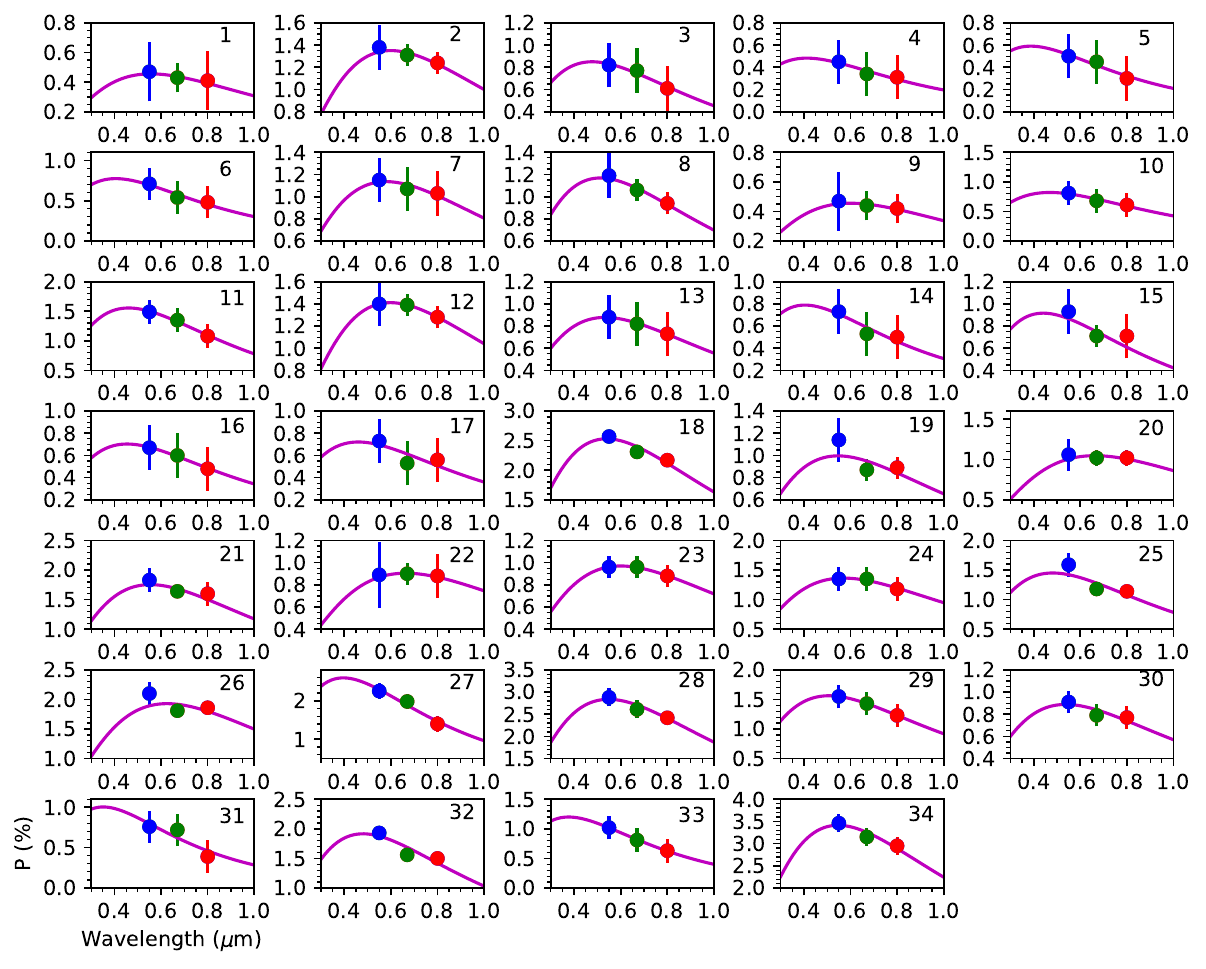}
\caption{Serkowski Curve for 34 bright Tycho stars observed in VRI filters. Blue, green, and red-filled circles correspond to V-band, R-band, and I-band, respectively. Magenta solid line represents the fitting using Serkowski Law \citep{1975ApJ...196..261S}.}\label{fig:Serkowski}
\end{figure*}

The Serkowski law \citep{1975ApJ...196..261S} with the parameter K provides an appropriate description of the interstellar polarization with the wavelength in the visible region, and it is defined as follows:

\begin{equation}\label{eq:Serkowski}
\mathrm{P_\lambda = P_{max}~exp[-K~ln^2(\lambda_{max}/\lambda)]}
\end{equation}

where $\mathrm{P_{\lambda}}$ is the degree of polarization (\%) at wavelength $\lambda$, $\mathrm{P_{max}}$ is the maximum degree of polarization estimated from the fitting at wavelength $\lambda_{\mathrm{max}}$ and the parameter K determines the width of the peak in the curve and adopted as K = 1.15 \citep{1975ApJ...196..261S}. If the polarization is produced by aligned interstellar dust grains, the observed data will follow equation \ref{eq:Serkowski} and the values of $P_{\mathrm{max}}$ and $\lambda_{\mathrm{max}}$ for each star can be calculated. We studied the dust properties of LOSFR by obtaining the $P_{\mathrm{max}}$ and $\lambda_{\mathrm{max}}$ values using the weighted least-squares fitting to the measured polarization in V, R, and I bands.
\begin{table}
\centering
\caption{The mean values of $\lambda_\mathrm{{max}}$ and P$_\mathrm{{max}}$ for UBVRI, BVRI and VRI data set for various regions.}\label{tab: BVRI}
\begin{tabular}{c c c}\hline
Filter set & $\lambda_\mathrm{{max}}$ ($\mu$m)&  P$_\mathrm{{max}}$(\%) \\ \hline
\multicolumn{3}{c}{\bf NGC 5617}\\
UBVRI      & 0.53 $\pm$ 0.05       & 4.13 $\pm$ 0.22 \\
BVRI       & 0.57 $\pm$ 0.06       & 4.05 $\pm$ 0.21 \\
VRI        & 0.61 $\pm$ 0.05       & 3.98 $\pm$ 0.16  \\ \hline
\multicolumn{3}{c}{\bf NGC 5749}\\
UBVRI      &  0.60 $\pm$ 0.05      & 1.77 $\pm$ 0.08 \\
BVRI       &  0.62 $\pm$ 0.09      & 1.76 $\pm$ 0.09 \\
VRI        &  0.65 $\pm$ 0.09      & 1.75 $\pm$ 0.10 \\ \hline
\multicolumn{3}{c}{\bf LDN 1570}\\
BVRI       &  0.63 $\pm$ 0.05      & 2.86 $\pm$ 0.08 \\
VRI        &  0.63 $\pm$ 0.05      & 2.86 $\pm$ 0.08 \\ \hline
\multicolumn{3}{c}{\bf NGC 457}\\
BVRI       &  0.57 $\pm$ 0.03      & 2.96 $\pm$ 0.09 \\
VRI        &  0.56 $\pm$ 0.04      & 2.97 $\pm$ 0.15 \\ \hline
\end{tabular}
\end{table}

The $\lambda_{\mathrm{max}}$ values can also be used to infer the origin of the polarization. The stars with $\lambda_{\mathrm{max}}$ much lower than the average value of the interstellar medium \citep[0.55 $\mu$m;][]{1975ApJ...196..261S} may have an intrinsic component of polarization due to scattering process \citep{OrsattiVegaMarraco1998}. Fig. \ref{fig:Serkowski} shows the Serkwoski fitting for 34 stars where we obtain the $\lambda_{max}$ values between 0.37 $\mu$m to 0.66 $\mu$m. The weighted mean values of the $P_{\mathrm{max}}$ and $\lambda_{\mathrm{max}}$ for 34 stars are found to be 1.30 $\pm$ 0.11 $\%$ and 0.51 $\pm$ 0.05 $\mu$m, respectively. Since we have made observations only at three wavelength bands (VRI), we have investigated the effect of using only three wavelength bands on the measured $P_{\mathrm{max}}$ and $\lambda_{\mathrm{max}}$. Toward this, we have compiled the polarimetric data of some randomly selected clouds, e.g., NGC 5617, NGC 5749 in UBVRI and LDN 1570, NGC 457 in BVRI filters from \cite{2010A&A...513A..75O}, \cite{2007A&A...462..621V}, \cite{2013A&A...556A..65E} and \cite{2017PASP..129j4201T}, respectively and estimated the values of $P_{\mathrm{max}}$ and $\lambda_{\mathrm{max}}$ by reducing the number of wavelength bands i.e., using the UBVRI, BVRI, and VRI filters. The mean values of $\lambda_{\mathrm{max}}$ and $P_{\mathrm{max}}$ for these three regions are shown in Table \ref{tab: BVRI}. Using BVRI data set, the mean values of $\lambda_{max}$ and P$_{max}$ for NGC 5749 are found to be 0.60 $\pm$ 0.05 $\mu$m and 1.61 $\pm$ 0.08\% while using only VRI filters the values are found to be 0.69 $\pm$ 0.07 $\mu$m and 1.59 $\pm$ 0.07\%. Similarly, The mean values of $\lambda_{max}$ and P$_{max}$ for LDN 1570 are found to be 0.61 $\pm$ 0.04 $\mu$m and 2.90 $\pm$ 0.11\% from BVRI data set and 0.63 $\pm$ 0.07 $\mu$m and 2.85 $\pm$ 0.15\% from VRI data set. We found that the values of $\lambda_{max}$ for the VRI data set are consistent with the uncertainty compared with UBVRI and BVRI data sets. The difference in $\lambda_\mathrm{{max}}$ and $P_{\mathrm{max}}$ values in different filter sets is within 10\%. Hence, our results can be used to estimate the dust grain size in LOSFR. The estimated $\lambda_\mathrm{{max}}$ is 0.51 $\pm$ 0.05 $\mu$m, which is consistent with the value corresponding to the general interstellar medium \citep[0.55 $\mu$m;][]{1975ApJ...196..261S}. It is also consistent with a value 0.53 $\mu$m for LOSFR along the Galactic plane estimated from the empirical relation, $\lambda_\mathrm{{max}}$ = 0.545 + 0.030 sin({\textit{l}} + 175$\degree$) \citep{Whittet1977}. This relation describes the systematic modulation of $\lambda_\mathrm{{max}}$ as a function of Galactic latitude ({\textit{l}}). We also estimated the value of R$_{V}$, the total-to-selective extinction, using the relation R$_{V}$ = (5.6 $\pm$ 0.3) $\times$ $\lambda_\mathrm{{max}}$ \citep{1978A&A....66...57W}, for all the 34 stars. The R$_{V}$ values range from 1.96 to 3.70 for all 34 Tycho stars. The mean value of R$_{V}$ for these 34 stars comes out to be 2.9 $\pm$ 0.3, which is in agreement with the general average value (R$_{V}$ = 3.1) for the Milky Way Galaxy within error limits, indicating that the size of the dust grains within the LOSFR is normal.

\subsection{Magnetic field geometry of BRC 17 and BRC 18}\label{ch4:BRC17-18}

\begin{figure*}[h]
\centering
\includegraphics[width=8.8cm]{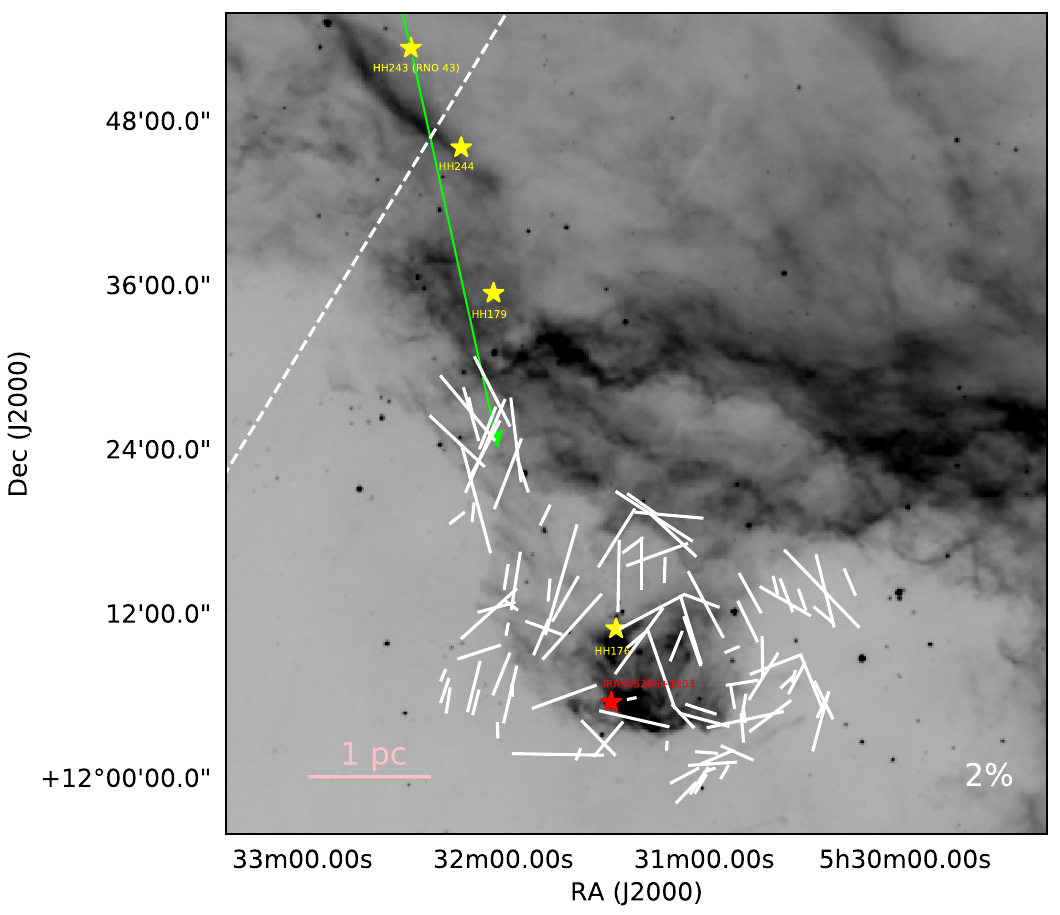}
\includegraphics[width=9.2cm]{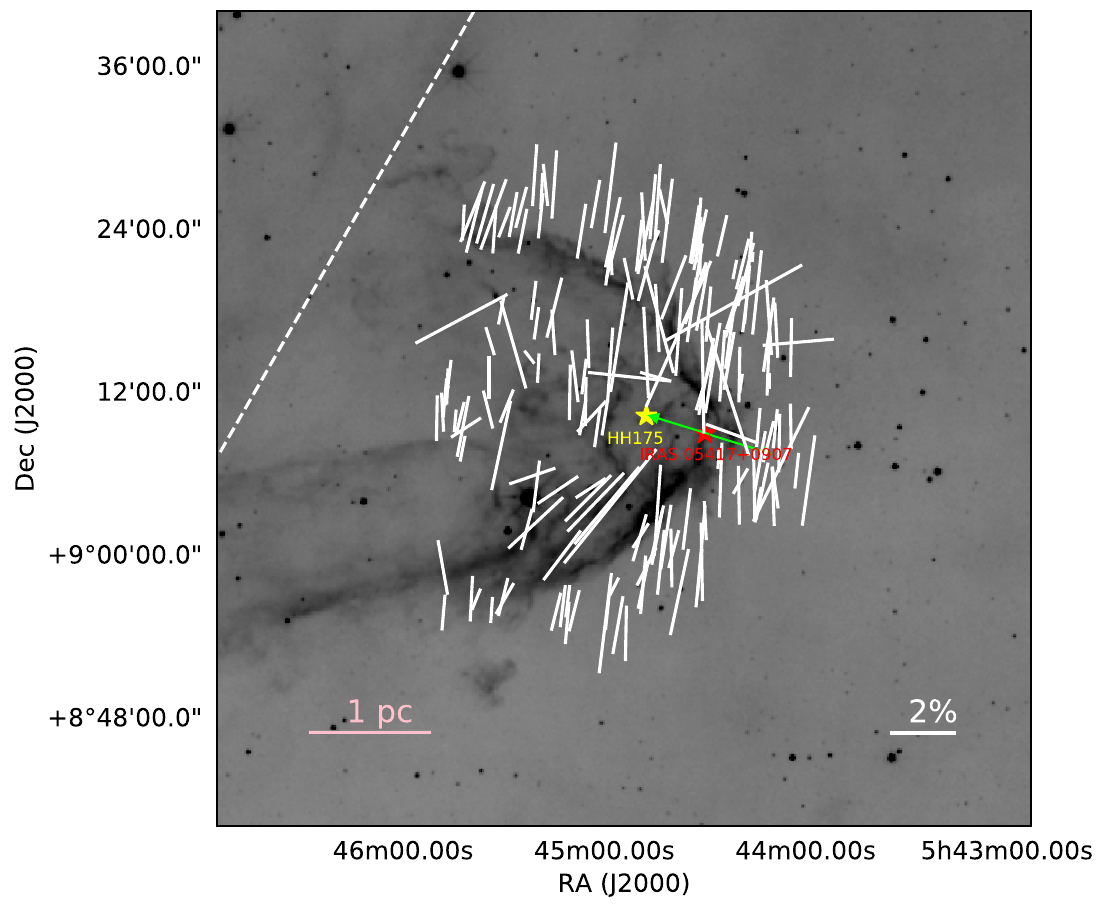}
\caption{Polarization vectors are over-plotted on WISE 12$\mu$m image of BRC 17 (left panel) and BRC 18 (right panel). The length of the vectors corresponds to the degree of polarization, and orientation corresponds to the position angle measured from the north and increasing toward the east. A vector corresponding to 2$\%$ polarization is shown for reference. The white dashed line shows the orientation of the Galactic plane. The red star shows the location of IRAS sources, and the yellow stars show the positions of HH objects in both the images \citep{2000yCat.5104....0R}. Green arrows in both panels show the directions of outflows. The outflow sources are HH 243 for BRC 17, while IRAS 05417+0907 for BRC 18.} \label{fig:SFO17_P}
\end{figure*}

The magnetic field geometry of a molecular cloud is predominantly regulated by the relative dynamical importance of magnetic forces to gravity, turbulence, and thermal pressure. If the magnetic field lines strongly support the molecular cloud against gravity, then the magnetic field lines would be aligned smoothly. The average magnetic field orientation would be perpendicular to the major axis of the cloud \citep{Mouschovias1978}. It is believed that the cloud tends to shrink more in the parallel direction to the magnetic field rather than the perpendicular direction. However, if the turbulence is more important than the magnetic field, then the structural dynamics of the molecular cloud would be controlled by the random motions, and the turbulent eddies will drag the magnetic field lines around \citep{BallesterosParedes1999b}. Thus, the magnetic field lines would become chaotic with no preferred direction. 

The results of our optical polarimetry observations of 105 stars in the direction of BRC 17 and 171 stars towards BRC 18 are presented in Table \ref{tab:Polresult}. We have tabulated the results of only those sources having a degree of polarization measurements better than 2$\sigma$. The columns of these tables show the star identification number in increasing order of their right ascension, the declination, measured P ($\%$), and polarization position angles ($\theta$ in degrees). 

\subsubsection{BRC 17}

\begin{figure*}
\sidecaption
  \includegraphics[width=6.2cm]{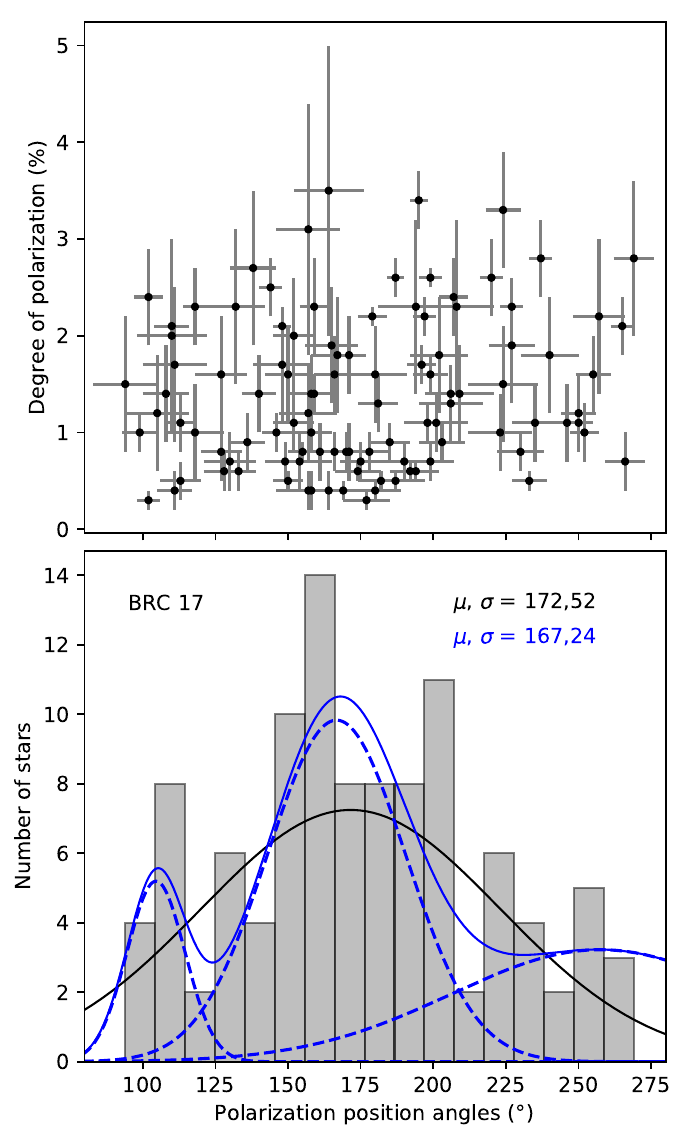}
  \includegraphics[width=6.2cm]{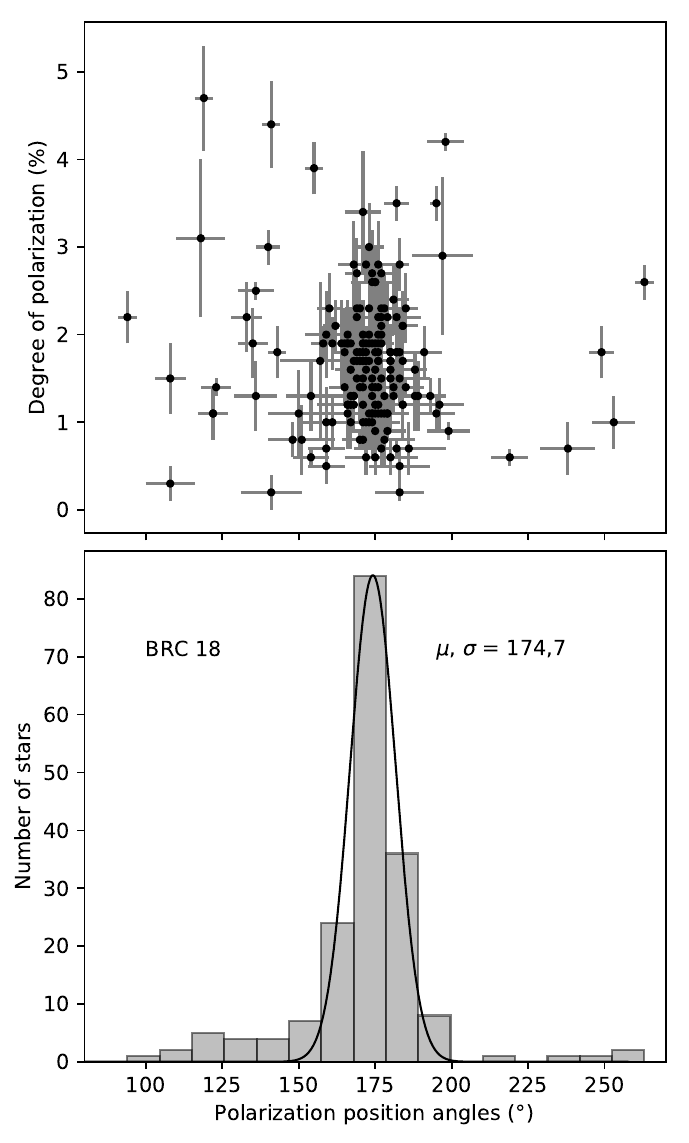}
     \caption{Degree of polarization versus the polarization position angle of stars projected on BRC 17 (left panel) and BRC 18 (right panel). The histogram of the position angles is also shown. These values are obtained after subtracting the foreground interstellar component from the observed values. The solid black curves represent a fit to the histogram of position angles using a single Gaussian model. BRC 17 is also fitted with a multiple Gaussian model (solid-blue) where decomposed Gaussians are shown by dashed-blue along with the mean and standard deviation of the central Gaussian marked.}\label{fig:hisogram}
\end{figure*}

The cloud complex B30 contains Barnard clouds B31 and B32, and the Lynds clouds LDN1580-1584 \citep{2001AJ....121.2124D}. At the head part of this cloud, an optically visible bright rim is present, and this is cataloged as BRC 17 by \citealt{1991ApJS...77...59S}. The distance of BRC 17 is estimated at around 389 $\pm$ 9 pc (see section \ref{distance}). The projected angular distance of the BRC 17 from the ionizing star $\lambda-$Ori is $\sim$2.3$\degree$, which corresponds to $\sim$15 pc considering the cloud distance 389 pc. \citealp{1988ApJ...333..809Z} mapped this cloud using CO observations and divided it into two parts: The northern part containing RNO 43 outflow and the Southern part containing bright-rim and IRAS source. We mapped the magnetic field geometry towards the Southern region, which contains the bright rim of this cloud. The left panel of Fig. \ref{fig:SFO17_P} shows the plot of polarization vectors of 105 foreground-subtracted stars overplotted on the WISE 12 $\mu$m image. The length and the orientation of the vectors correspond to the measured P$\%$ and $\theta$ values, respectively. The $\theta$ is measured from the north, increasing toward the east. A vector with 2\% polarization is shown for reference. The dashed white line represents the orientation of the Galactic plane at the Galactic latitude of $-11.6\degree$. The mean values of the degree of polarization and polarization angle for BRC 17 are 1.4 $\pm$ 0.4$\%$ and 175 $\pm$ 43$\degree$, respectively. The plot of P\% vs. $\theta$ and the histogram of the $\theta$ after correcting for the foreground interstellar contribution are shown in Fig. \ref{fig:hisogram}. At first, we performed a single Gaussian fit to the histogram with bins size of 10 deg. The mean and standard deviation of the best-fit Gaussian is found to be $172\pm 9$ deg and $52\pm 12$ deg, respectively. The distribution of polarization position angle appears to be a mixture of multiple Gaussians with the strongest central peak. Therefore, we also fitted a multiple Gaussian model consisting of three Gaussians to the data. The multi-Gaussian model shows the strongest peak at the center with a mean and standard deviation of $167\pm 14$ deg and $24\pm 19$ deg, respectively. Although the different peaks in the multi-Gaussian model are not well separated and parameters are not well constrained, having large uncertainties, we find a similar reduced $\chi^2$ as that of the single-Gaussian model. However, the standard deviation value of the central Gaussian in the multi-Gaussian model is much smaller than the single Gaussian model.

\subsubsection{BRC 18} \label{ch4:brc18}
BRC 18 (also known as B35) lies at the periphery of the LOSFR. It is located at $\sim$392 $\pm$ 7 pc (see section \ref{distance}) with a dimension of 0.37 pc by 1.40 pc. The projected angular distance between the BRC 18 and the ionizing star $\lambda-$Ori is $\sim$2.5$\degree$, corresponding to $\sim$17 pc considering the cloud distance 392 pc. The magnetic field geometry in BRC 18 is found to be regular and oriented in the perpendicular direction of the cloud's major axis (along the east-west direction in the right panel of Fig. \ref{fig:SFO17_P}), indicating the magnetic field dominance in the cloud \citep{Mouschovias1978}. The mean values of P and $\theta$ for BRC 18 are 1.7 $\pm$ 0.3$\%$ and 171 $\pm$ 8$\degree$, respectively. The right panel of Fig. \ref{fig:SFO17_P} shows the plot of polarization vectors of 171 foreground-subtracted stars overplotted on the WISE 12 $\mu$m image. As discussed in the previous section, the length and orientation of the polarization vectors correspond to the measured P$\%$ and $\theta$ values, respectively. The $\theta$ is measured from the north, increasing toward the east. A vector with 2\% polarization is shown for reference. The white-dashed line represents the orientation of the Galactic plane at the galactic latitude of $-10.4\degree$. The plot of P\% vs. $\theta$ and the histogram of the $\theta$ with a bin size of 10 deg after correcting for the foreground interstellar contribution are shown in Fig. \ref{fig:hisogram}. The yellow star corresponds to the HH object, HH 175 \citep{2000yCat.5104....0R}. In BRC 18, the relatively low dispersion ($\sigma_{\theta}$=8$\degree$) in polarization position angles implies that the magnetic field lines are well aligned (see right panel in Fig. \ref{fig:SFO17_P}). The mean polarization position angle, $\sim$171$\degree$, from the north towards the east shows the direction of the plane-of-the-sky component of the magnetic field, whereas the projected orientation of incoming ionizing radiation from $\lambda-$Ori towards BRC 18 $\sim$109$\degree$ for the north increasing towards the east. Hence, the projected angle between the magnetic field lines and the incoming ionizing radiation is $\sim$62$\degree$. This suggests the magnetic field lines in BRC 18 are $\sim$30$\degree$ off to the perpendicular direction to the incoming ionizing radiation. 

\subsection{Magnetic field strength using structure function analysis}\label{B_strength}

After knowing the direction of the magnetic field, we estimated the magnetic field strength using the improved Davis–Chandrasekhar–Fermi method (DCF; \citealt{DavisGreenstein1951, 1953ApJ...118..113C}) by determining the turbulent angular dispersion \citep{2009ApJ...696..567H, 2009ApJ...706.1504H}. In this method, the dispersion of the polarization angles as a function of distance or Angular Dispersion Function (ADF) is estimated by considering that the net magnetic field consists of a large-scale structured field, B$_\mathrm{0}$(x), and a turbulent component, B$_\mathrm{t}$(x). The angular dispersion function (ADF) or the square root of the structure-function (SF) is defined as the root mean squared differences between the polarization angles measured for two points separated by a distance \emph{l} and measured as shown in eq \ref{eq1}. Hence, SF analysis can be used as an important statistical tool to correlate the large-scale structured field and the turbulent component of the magnetic field in molecular clouds \citep[e.g., ][]{2009ApJ...696..567H, 2010ApJ...723..146F, 2012ApJ...751..138S, 2013A&A...556A..65E, 2016A&A...588A..45N}.

\begin{equation}
\langle\Delta\phi^{2}(l)\rangle^{1/2} =\Bigg \{\frac{1}{N(l)} \sum\limits_{i = 1}^{N(l)} [\phi(x) - \phi(x+l)]^{2} \Bigg \}^{1/2} \label{eq1}
\end{equation}

The structure function within the range $\delta < \emph{l} \ll d$, (where $\delta$ and $d$ are the correlation lengths which characterize B$_\mathrm{t}$(x) and B$_\mathrm{0}$(x), respectively), can be estimated using eq \ref{eq:2}:
\begin{equation}
\langle\Delta\phi^{2}(l)\rangle_{\mathrm{tot}} \simeq b^{2} + m^{2}l^{2} + \sigma^{2}_{\mathrm{M}}(l)\label{eq:2}
\end{equation} 
where $\langle\Delta\phi^{2}(l)\rangle_{\mathrm{tot}}$ shows the total measured dispersion estimated from the data. $\sigma^{2}_{\mathrm{M}}(l)$ are the measurement uncertainties and are calculated by taking the mean of the variances on $\Delta\phi(l)$ in each bin. The quantity $b^{2}$ is the constant turbulent contribution, estimated by the intercept of the fit to the data after subtracting $\sigma^{2}_{\mathrm{M}}(l)$. The term $m^{2}l^{2}$ is a smoothly increasing contribution with the length $l$ ($m$ shows the slope of this linear behavior). All these quantities are statistically independent from each other.

The ratio of the turbulent component and the large-scale magnetic fields is calculated by the eq \ref{eq:3}:
\begin{equation}
\frac{\langle B^{2}_\mathrm{t}\rangle^{1/2}}{B_\mathrm{0}} = \frac{b}{\sqrt{2 - b^{2}}}\label{eq:3}
\end{equation}

\begin{figure}
  \resizebox{\hsize}{!}{\includegraphics{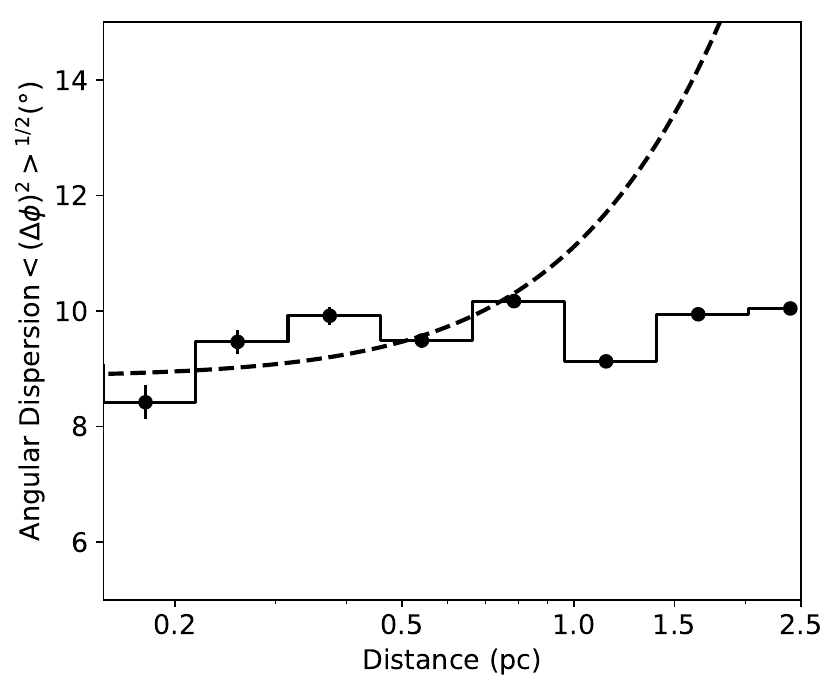}}
  \resizebox{\hsize}{!}{\includegraphics{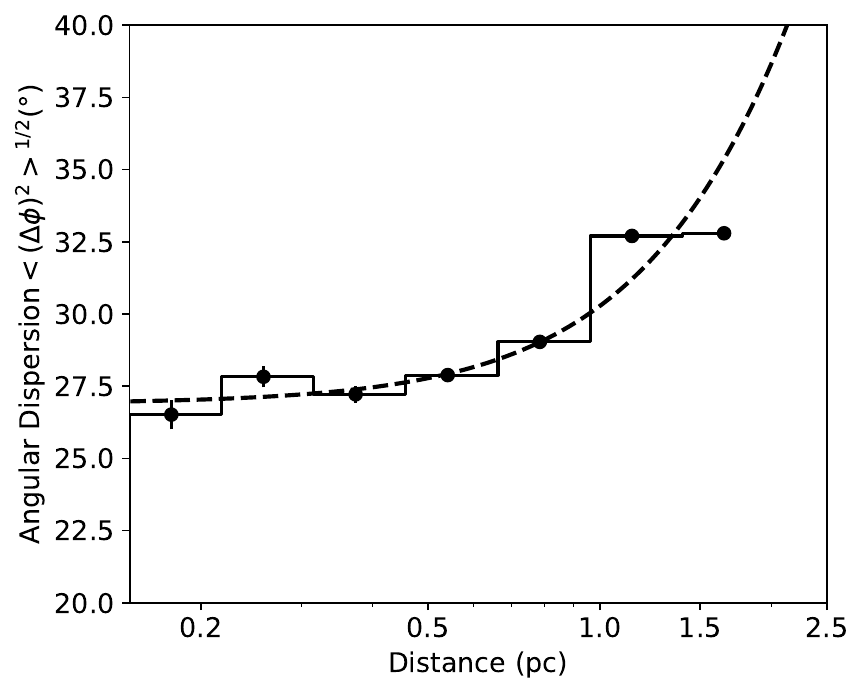}}
  \caption{Angular dispersion function (ADF) of the polarization angles, $\langle\Delta\phi^{2}(l)\rangle^{1/2}$ (\degree), with distance (pc) for BRC 18 (upper panel) and BRC 17 (lower panel). The dashed line denotes the best fit to the data up to 1pc distance.}
  \label{fig:structureFunc_BRC18}
\end{figure}

For the BRC 18 region, we estimated the ADF and plotted with the distance in Fig. \ref{fig:structureFunc_BRC18}. We used the polarization angle of 171 stars for BRC 18 to calculate ADF. The measured errors in each bin are very small, which are also over-plotted in Fig. \ref{fig:structureFunc_BRC18}. Each bin denotes the $\sqrt{\langle\Delta\phi^{2}(l)\rangle_{\mathrm{tot}} - \sigma^{2}_{\mathrm{M}}(l)}$ which is the ADF corrected for the measurement uncertainties. Bin widths are taken on a logarithmic scale. We used only 5 points, corresponding to a distance less than 1 pc, of the ADF in the linear fit of eq \ref{eq:2} shown by the dashed line. The shortest distance we considered is $\simeq$ 0.15 pc. The net turbulent contribution to the angular dispersion, $b$, is calculated to be 8.9$\degree$ $\pm$ 0.3$\degree$ (0.15 $\pm$ 0.005 rad) for BRC 18. Then, we estimated the ratio of the turbulent component and the large-scale magnetic fields using eq \ref{eq:3}, which is found to be 0.08 $\pm$ 0.004 for BRC 18. The result shows that the turbulent component of the magnetic field is very small compared to the large-scale structured magnetic field, i.e., $B_\mathrm{t} \ll B_\mathrm{0}$.

The strength of the plane of the sky component of the magnetic field was estimated from modified CF relation \citep{2015ApJ...807....5F} as shown in eq \ref{equ:B_field},
\begin{equation}
B_\mathrm{pos} = 9.3 \left[ \frac{2~n(H_{2})}{cm^{-3}} \right]^{1/2} \left[ \frac{\Delta V}{km s^{-1}} \right] \left[\frac{b}{1^{\circ}} \right]^{-1}~~\mu G \label{equ:B_field}
\end{equation}

In eq \ref{equ:B_field}, $n(H_{2})$ is the volume hydrogen density of the molecular hydrogen. It is obtained by estimating the hydrogen column density of the region probed by our optical polarimetry and from the radius of the cloud, assuming it to be a spherical cloud. We used the relation, N(H$_\mathrm{2}$)/$A_\mathrm{V}$ = 9.4 $\times 10^{20}$ cm$^{-2} mag^{-1}$ \citep{1978ApJ...224..132B} for estimating the column density. Based on the method described in the papers \citet{2010A&A...509A..44M} and \citet{2016A&A...588A..45N}, the average value of extinction traced by the stars lying behind the cloud (assuming distance $\gtrsim$ 400 pc for BRC 18) observed in this study is found to be $\sim0.6$ mag for BRC 18. We also determined the visual extinction, $A_\mathrm{V}$ using the near-infrared color excess revised (NICER) extinction maps that are based on the 2MASS data \citep{2016A&A...585A..38J}. The value of $A_\mathrm{V}$ is found to be 1.6 mag for BRC 18. This is because the method is only sensitive to the extinction of those stars lying on the cloud periphery. Hence, we have used the average value of $A_\mathrm{V}$, which is found to be 1.1 mag. We estimated the radius of BRC 18 by fitting a circle on the WISE 12 $\mu$m image. The radius of BRC 18 is determined as $\sim$1.5 pc, and hence volume density is found to be $\sim$230 cm$^{-3}$ for BRC 18. We adopted the $^{12}$CO line width, $\Delta V$ = 1.7 km s$^{-1}$ for BRC 18, from our molecular line observations (Neha et al.; in preparation). After substituting all the values in eq \ref{equ:B_field}, the value of $B_\mathrm{pos}$ is found to be $\sim$40 $\mu$G for BRC 18. In this way, the BRC 18 cloud has a magnetic field strength $\sim$40 $\mu$G, comparable to the moderate magnetic field strength of the simulations, in an almost perpendicular direction to the ionizing radiation. On comparing this with the simulations, we found that the photoionization shock can be passed through the cloud without any hindrance and can trigger the star formation \citep{2011MNRAS.414.1747A}. 

Polarization position angles for BRC 17, as shown in Fig. \ref{fig:hisogram}, exhibit multiple Gaussian features. We used 71 out of 105 sources whose polarization angles follow the central component in the multi-Gaussian model to estimate the magnetic field strength since it shows a relatively smaller dispersion than the whole data and follows the ambient magnetic field direction. We performed structure-function analysis as described above. The lower panel of Fig. 7 shows a plot of the ADF as a function of distance for BRC 17. Similar to BRC 18, we used 5 points of ADF and performed a linear fit using eq. 3. The value of b is $26.9\degree \pm 0.3\degree$ for BRC 17. The ratio of turbulent to the large-scale component of the magnetic field is found to be $0.35\pm0.22$, suggesting the turbulent component is smaller than the large-scale structured magnetic field. We then estimated the strength of the plane of the sky component of the magnetic field using eq. 5, adapting n($H_2$) = 2000 cm$^{-3}$ and $\Delta V= 1.3$ km s$^{-1}$ from \citet{1988ApJ...333..809Z}, who performed molecular line observation of BRC 17. The magnetic field is $\sim$28 $\mu$G, which can be considered weak magnetic field strength. Note that the calculated value of $B_\mathrm{pos}$ should be considered a rough estimate due to the large uncertainties of the individual quantities involved in its calculation, particularly for BRC 17, which has a larger b value.

\section{Discussion}
Based on the radiation-magnetohydrodynamics (R-MHD) simulations of HII regions, \cite{2011MNRAS.414.1747A} presented the large-scale magnetic field maps projected on the whole HII region, and they showed that the magnetic field lines are mainly oriented along the large-scale ionization fronts, that forms a ring around the HII region. The expanding HII and photo-dissociation regions (PDR) remove the pre-existing small-scale disordered magnetic field pattern and produce a large-scale ordered magnetic field in the neutral shell, with approximately parallel orientation to the ionization front. They also presented that the magnetic field lines are aligned in a perpendicular direction to the ionization front at the head of the globules associated with the HII region. Furthermore, \cite{2009MNRAS.398..157H} and \cite{2011MNRAS.412.2079M} included the effect of magnetic field on the globules in the presence of ionizing radiation from the OB star. \cite{2011MNRAS.412.2079M} performed 3D-radiation magneto-hydrodynamics (3D-RMHD) simulations considering three strengths of magnetic fields, 18 $\mu$G (weak), 53 $\mu$G (medium) and 160 $\mu$G (strong) for perpendicular direction to the ionizing radiation. They found that the RDI process significantly changes the initially perpendicular weak field orientation. The weak field is aligned along the incoming ionizing radiation due to the RDI process and rocket effect. In the medium perpendicular case, slight changes were noticed in the field direction, whereas the strong perpendicular magnetic field remains unchanged. Hence, the strong magnetic field can alter the morphology of the structure developed due to RDI and rocket effect, partially due to shielding by the dense ionized ridge and partially due to the effect of the magnetic field within the globule.

\begin{figure}
\centering
\resizebox{8.5cm}{7.5cm}{\includegraphics{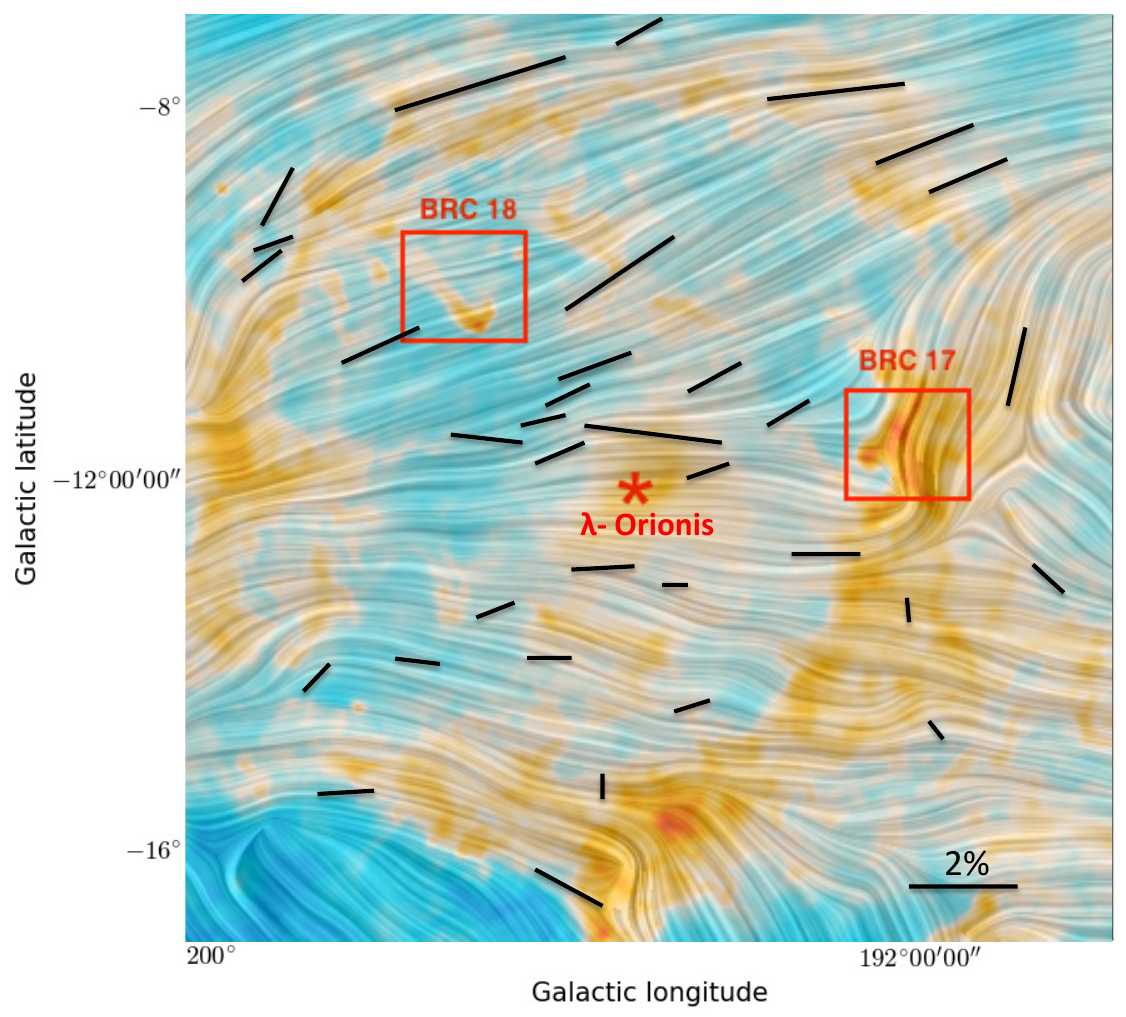}}
\caption{\small Global magnetic field morphology in LOSFR. The layered texture in the image represents the magnetic field geometry using sub-mm polarization, while the black vectors, overplotted on the Planck sub-mm polarization map, correspond to the optical magnetic field geometry using observed R-band polarization. The red star shows the position of $\lambda$-Ori star. The clouds, BRC 17 and BRC 18 are identified. A 2\% degree of polarization reference vector has been shown for optical polarization. The image is taken by ESA's Planck satellite (image credit: ESA and the Planck Collaboration).} \label{fig:planck}
\end{figure}

\begin{figure*}
\centering
   \includegraphics[width=9.1cm]{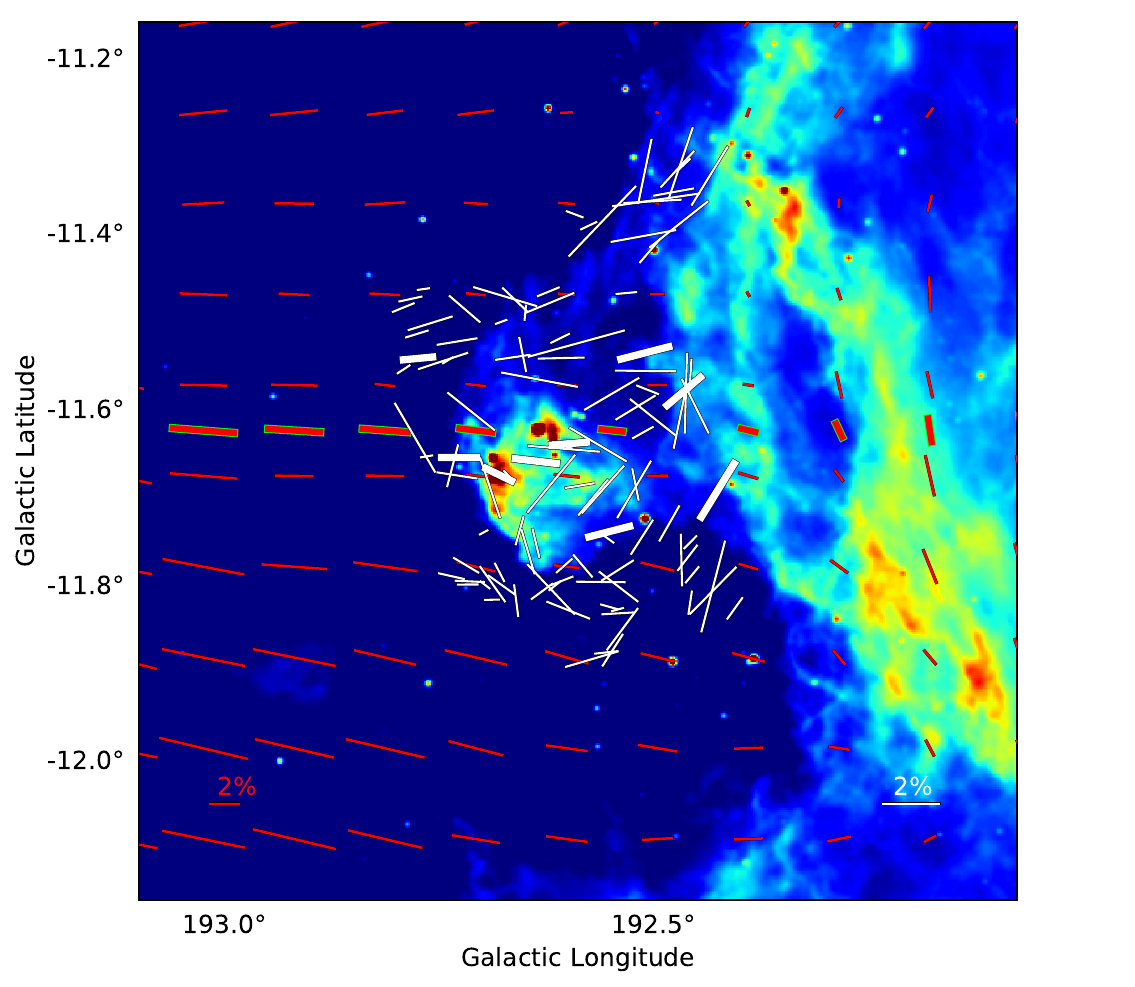}
   \includegraphics[width=9.1cm]{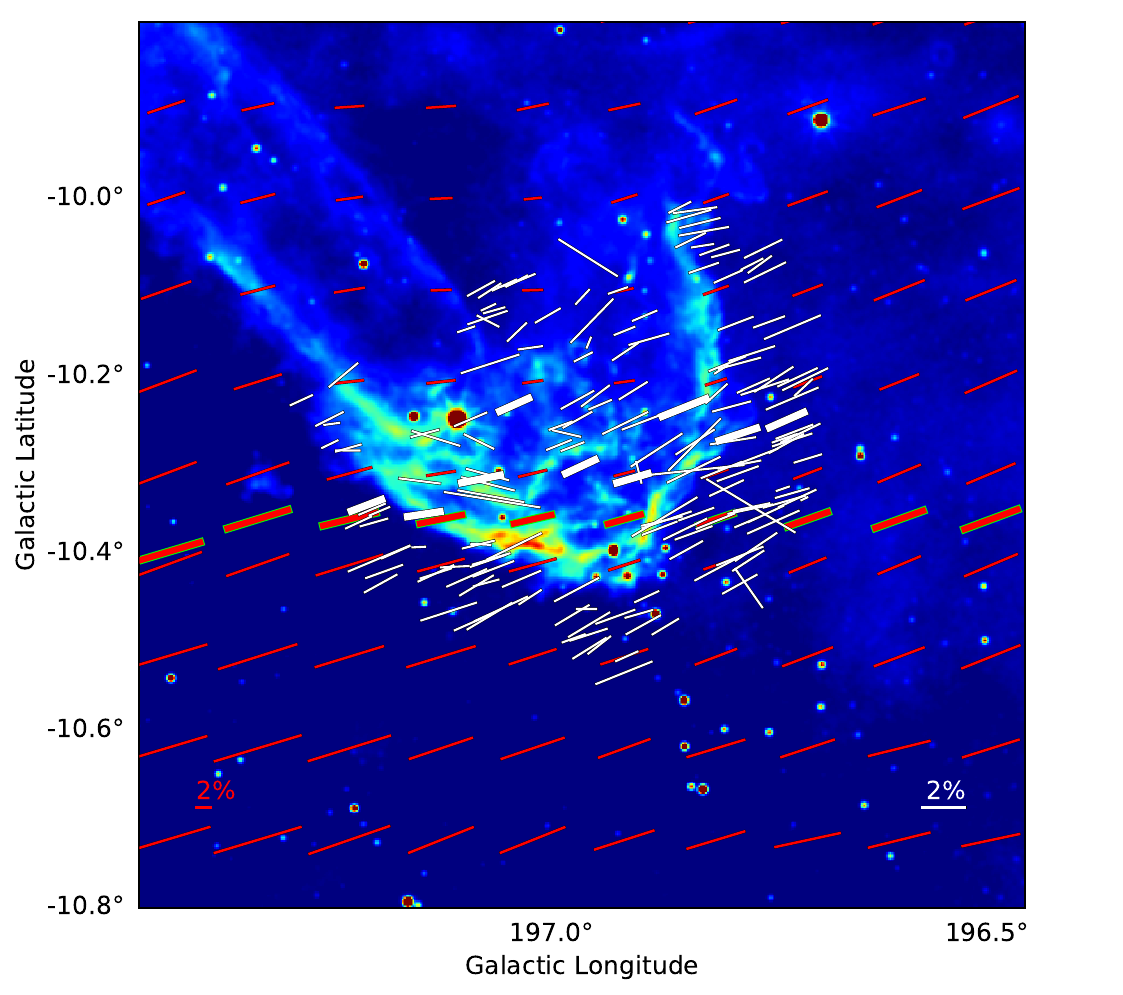}
     \caption{Polarization vectors are over-plotted on WISE 12$\mu$m image of BRC 17 (left panel) and BRC 18 (right panel) of our optical data (white color) and of Planck data (red color). Binned Polarization vectors along galactic longitude bins are shown in both panels for our data (thick white) and Planck (thick red). The length of the vectors corresponds to the degree of polarization, and orientation corresponds to the position angle measured from the north and increasing toward the east. Vectors corresponding to 2$\%$ polarization are shown for references in their respective colors.}
     \label{fig: SFO17_18_planck}
\end{figure*}

The direction of polarization vectors represents the global plane-of-sky magnetic field geometry in LOSFR, which seems to follow the large-scale structure seen in all the images of Fig. \ref{fig:V_band}, which is consistent with the simulations shown by \cite{2011MNRAS.414.1747A}. Though our sample is small, the polarization vectors follow the ring of the LOSFR clearly in Fig. \ref{fig:V_band}. Fig. \ref{fig:planck} shows the global galactic magnetic field geometry in LOSFR in optical and sub-mm wavelengths. The Planck image,\footnote{\url{http://sci.esa.int/planck/55910-polarised-emission-from-orion/}} taken by ESA's Planck satellite (\citealt{2016A&A...594A...1P}), shows the sub-mm polarization map indicating the global magnetic field morphology. In contrast, the R-band polarization vectors in black color, corresponding to the optical magnetic field geometry, are overplotted. In the upper part of the image, the magnetic field geometry seems to be organized and regular, which could be due to the large-scale orientation of the magnetic field lines along the Galactic plane. Our optical polarization data is consistent with the sub-mm, and the orientation of the magnetic field follows the same pattern. It is also seen locally around BRC 17 and BRC 18 as shown in Fig. \ref{fig: SFO17_18_planck}, where the Planck sub-mm polarization vectors along with optical polarization vectors towards BRC 17 and BRC 18 are overplotted on WISE 12 $\mu$m image. These Planck polarization vectors were extracted using Planck public data release 3 (\citealt{2016A&A...594A...1P}). Since the polarization techniques for both optical and sub-mm wavelengths are different, as the polarization occurs in optical due to partial extinction while in sub-mm wavelengths, the polarization occurs due to emission, comparing the polarization maps of both wavelengths would be interesting.

Polarization observations of BRC 17 show a large range of polarization angles with a central peak at 167\degree and two additional components far away. This could be due to the combined effects of a shock wave coming from central ionizing source $\lambda-$Ori and the presence of HH objects in the vicinity. An abrupt change in the direction of the magnetic field around BRC 17 is noticed, consistent with the Planck observations as shown in Fig. \ref{fig: SFO17_18_planck}. We have binned polarization vectors along the galactic longitude for a clear view. Although we lack observations below the galactic longitude of 192.5\degree, our polarization vectors show a similar direction to that of Plank, being consistent above longitude >192.5\degree and turning towards the high-density material following Planck vectors. This magnetic field orientation could be the reason for the accumulation of high column densities there, as clouds generally tend to form at ``kinks" or ``bends" in the magnetic field (\citealt{1995ApJ...455..536P}; \citealt{2001ApJ...562..852H}; \citealt{2001ApJ...547..280H}; \citealt{2001ApJ...546..980O}). 

\citet{2004Ap.....47..162M} searched HH-objects and emission line stars in star-forming regions and observed an intense molecular outflow spread over $\sim$5 pc. This is one of the high-velocity and largest giant outflows known to date. This outflow is known as RNO 43 associated with the source (HH 243) IRAS 05295+1247 \citep{1996MNRAS.279..866B}, having the outflow velocity around 11 km s$^{-1}$ \citep{2004A&A...426..503W}. Few more HH objects are present in the BRC 17 region. At the head part of BRC 17, HH 176 is present, which is powered by a peculiar A4 type variable star HK-Ori \citep{2000A&A...357.1001L}. The presence of such a big high-velocity outflow and HH objects in the BRC 17 region suggest that the star formation activities are going on, and cloud material is dynamically more active. 
\cite{1988ApJ...333..809Z} performed molecular line observations towards BRC 17 and showed gas temperature and column density enhancements along the edge of the cloud, which corresponds to the optical bright rim. They also suggested that the gas-heating mechanism in the BRC 17 is most likely due to the photoelectric ejection of energetic electrons from dust grains induced by the far-ultraviolet radiation from the $\lambda-$Orionis OB association. However, these observations have very poor resolution. Further high-resolution molecular line observations can provide a better picture of the kinematics of BRC 17. We have calculated the magnetic field strength for BRC 17. The strength of the turbulent component is about 30\% of the large-scale magnetic field. The strength of the plane of sky component of the magnetic field is $\sim$28 $\mu$G, which can be considered in the weak field regime. Hence, it could be possible that the high-energy radiation is affecting the orientation of the magnetic field in BRC 17 \citep{2009MNRAS.398..157H, 2011MNRAS.412.2079M}. A detailed high-resolution polarimetric study is required to better understand the magnetic field geometry in BRC 17.

BRC 18 shows that the magnetic field lines are aligned in a preferred direction roughly perpendicular to the ionizing radiation, which is also consistent with the simulations \citep{2011MNRAS.414.1747A}. The directions of optical polarization vectors are also consistent with Planck as shown in Fig. \ref{fig: SFO17_18_planck}, where the binned polarization vectors along the galactic longitude are also plotted, similar to BRC 17. There is also one HH object, HH 175, present in the BRC 18 for which the driving source is IRAS 05417+0907, multiple systems with at least 6 components \citep{2021MNRAS.501.5938R}. Moreover, millimeter observations suggest that the BRC 18 cloud is found to have multiple cores, and the embedded source could be the result of compression by an ionization front driven by the central $\lambda-$Ori OB stars \citep{2021MNRAS.501.5938R}. We estimated the magnetic field strength for BRC 18 (see section \ref{B_strength}) and found $\sim$40 $\mu$G, a medium-strength magnetic field roughly perpendicular to the incoming ionization radiations. The moderate magnetic field strength does not affect the interaction of the ionization radiations with the cloud much but tries to align along the incoming ionizing radiations \citep{2011MNRAS.414.1747A}. Some of the polarization vectors in the lower region of BRC 18 (see right panel of Fig. \ref{fig:SFO17_P}) show that they align themselves along the incoming ionizing radiation, suggesting RDI in BRC 18.

\section{Conclusions} \label{conclusion}
We present optical polarimetric results for the stars globally projected on the LOSFR and the associated BRC 17 and BRC 18. Using these polarimetric observations, the importance of the magnetic field in the evolution of the HII region has been studied. We draw the following conclusions:

\begin{enumerate}

\item We estimated the distance of LOSFR as 392 $\pm$ 8 pc using astrometry information of YSOs towards BRC 17, BRC 18, and central cluster Collinder 69 members using GAIA EDR3.

\item  We found a large-scale ordered magnetic field in the LOSFR, and our results are consistent with the simulations and the Planck sub-mm polarization map. The orientation of the magnetic field is almost similar in all three optical bands, VRI, and sub-mm. 

\item The magnetic field lines traced by the polarization position angles in BRC 17 are roughly oriented in north-south directions but show a large dispersion. In the case of BRC 18, the magnetic field morphology is regular, having an approximately perpendicular orientation with the incoming ionizing radiations. The magnetic field geometry in these two regions is also consistent with the Planck magnetic field map.

\item The average dust size in the LOSFR is also determined using Serkowski's curve fitting, which comes out to be $\sim$0.51 $\pm$ 0.05 $\mu$m and the average value of total-to-selective extinction ratio, R$_{V}$ is equal to $\sim$2.9 $\pm$ 0.3 which is consistent with the general value of ISM for the Milky Way.

\item The turbulent to large scale magnetic field strength ratio is estimated to be higher $\sim$30 \% in BRC 17 compared to 10\% in BRC 18. The plane of the sky magnetic field strength, calculated using the structure-function analysis, is $\sim$28 $\mu$G for BRC 17 and $\sim$40 $\mu$G for BRC 18. 

\end{enumerate}

\begin{acknowledgements}
We thank the referee for her/his constructive suggestions/comments, which helped us to improve the quality of our manuscript. This research has made use of the SIMBAD database, operated at CDS, Strasbourg, France. We also acknowledge the use of NASA's \textit{SkyView} facility (http://skyview.gsfc.nasa.gov) located at NASA Goddard Space Flight Center. NS acknowledges the financial support provided under the National Post-doctoral Fellowship (NPDF; File Number: PDF/2022/001040) by the Science \&
Engineering Research Board (SERB), a statutory body of the Department of Science \& Technology (DST), Government of India. NS thanks Suvendu Rakshit for the useful discussions.
\end{acknowledgements}

\bibliographystyle{aa}
\bibliography{ref1}

\begin{thebibliography}{85}
\expandafter\ifx\csname natexlab\endcsname\relax\def\natexlab#1{#1}\fi

\bibitem[{{Arthur} {et~al.}(2011){Arthur}, {Henney}, {Mellema}, {de Colle}, \& {V{\'a}zquez-Semadeni}}]{2011MNRAS.414.1747A}
{Arthur}, S.~J., {Henney}, W.~J., {Mellema}, G., {de Colle}, F., \& {V{\'a}zquez-Semadeni}, E. 2011, \mnras, 414, 1747

\bibitem[{Bailer-Jones {et~al.}(2021)Bailer-Jones, Rybizki, Fouesneau, Demleitner, \& Andrae}]{Bailer_Jones_2021}
Bailer-Jones, C. A.~L., Rybizki, J., Fouesneau, M., Demleitner, M., \& Andrae, R. 2021, The Astronomical Journal, 161, 147

\bibitem[{{Ballesteros-Paredes} {et~al.}(1999){Ballesteros-Paredes}, {Hartmann}, \& {V{\'a}zquez-Semadeni}}]{BallesterosParedes1999b}
{Ballesteros-Paredes}, J., {Hartmann}, L., \& {V{\'a}zquez-Semadeni}, E. 1999, \apj, 527, 285

\bibitem[{{Barnard}(1919)}]{Barnard1919}
{Barnard}, E.~E. 1919, \apj, 49, 1

\bibitem[{{Bayo} {et~al.}(2012){Bayo}, {Barrado}, {Hu{\'e}lamo}, {Morales-Calder{\'o}n}, {Melo}, {Stauffer}, \& {Stelzer}}]{2012A&A...547A..80B}
{Bayo}, A., {Barrado}, D., {Hu{\'e}lamo}, N., {et~al.} 2012, \aap, 547, A80

\bibitem[{{Bence} {et~al.}(1996){Bence}, {Richer}, \& {Padman}}]{1996MNRAS.279..866B}
{Bence}, S.~J., {Richer}, J.~S., \& {Padman}, R. 1996, \mnras, 279, 866

\bibitem[{{Bertoldi}(1989)}]{1989ApJ...346..735B}
{Bertoldi}, F. 1989, \apj, 346, 735

\bibitem[{{Bohlin} {et~al.}(1978){Bohlin}, {Savage}, \& {Drake}}]{1978ApJ...224..132B}
{Bohlin}, R.~C., {Savage}, B.~D., \& {Drake}, J.~F. 1978, \apj, 224, 132

\bibitem[{{Chandrasekhar} \& {Fermi}(1953)}]{1953ApJ...118..113C}
{Chandrasekhar}, S. \& {Fermi}, E. 1953, \apj, 118, 113

\bibitem[{{Chauhan} {et~al.}(2009){Chauhan}, {Pandey}, {Ogura}, {Ojha}, {Bhatt}, {Ghosh}, \& {Rawat}}]{2009MNRAS.396..964C}
{Chauhan}, N., {Pandey}, A.~K., {Ogura}, K., {et~al.} 2009, \mnras, 396, 964

\bibitem[{Cortés {et~al.}(2021)Cortés, Sanhueza, Houde, Martín, Hull, Girart, Zhang, Fernandez-Lopez, Zapata, Stephens, bai Li, Wu, Olguin, Lu, Guzmán, \& Nakamura}]{Cortes_2021}
Cortés, P.~C., Sanhueza, P., Houde, M., {et~al.} 2021, The Astrophysical Journal, 923, 204

\bibitem[{{Cutri} {et~al.}(2003){Cutri}, {Skrutskie}, {van Dyk}, {Beichman}, {Carpenter}, {Chester}, {Cambresy}, {Evans}, {Fowler}, {Gizis}, {Howard}, {Huchra}, {Jarrett}, {Kopan}, {Kirkpatrick}, {Light}, {Marsh}, {McCallon}, {Schneider}, {Stiening}, {Sykes}, {Weinberg}, {Wheaton}, {Wheelock}, \& {Zacarias}}]{Cutrietal2003}
{Cutri}, R.~M., {Skrutskie}, M.~F., {van Dyk}, S., {et~al.} 2003, VizieR Online Data Catalog, 2246, 0

\bibitem[{{Davis} \& {Greenstein}(1951)}]{DavisGreenstein1951}
{Davis}, Jr., L. \& {Greenstein}, J.~L. 1951, \apj, 114, 206

\bibitem[{{Dolan} \& {Mathieu}(1999)}]{1999AJ....118.2409D}
{Dolan}, C.~J. \& {Mathieu}, R.~D. 1999, \aj, 118, 2409

\bibitem[{{Dolan} \& {Mathieu}(2001)}]{2001AJ....121.2124D}
{Dolan}, C.~J. \& {Mathieu}, R.~D. 2001, \aj, 121, 2124

\bibitem[{{Dolan} \& {Mathieu}(2002)}]{2002AJ....123..387D}
{Dolan}, C.~J. \& {Mathieu}, R.~D. 2002, \aj, 123, 387

\bibitem[{{Duerr} {et~al.}(1982){Duerr}, {Imhoff}, \& {Lada}}]{1982ApJ...261..135D}
{Duerr}, R., {Imhoff}, C.~L., \& {Lada}, C.~J. 1982, \apj, 261, 135

\bibitem[{{Elmegreen}(2011)}]{2011EAS....51...45E}
{Elmegreen}, B.~G. 2011, in EAS Publications Series, Vol.~51, EAS Publications Series, ed. C.~{Charbonnel} \& T.~{Montmerle}, 45--58

\bibitem[{{Eswaraiah} {et~al.}(2013){Eswaraiah}, {Maheswar}, {Pandey}, {Jose}, {Ramaprakash}, \& {Bhatt}}]{2013A&A...556A..65E}
{Eswaraiah}, C., {Maheswar}, G., {Pandey}, A.~K., {et~al.} 2013, \aap, 556, A65

\bibitem[{{Eswaraiah} {et~al.}(2012){Eswaraiah}, {Pandey}, {Maheswar}, {Chen}, {Ojha}, \& {Chandola}}]{2012MNRAS.419.2587E}
{Eswaraiah}, C., {Pandey}, A.~K., {Maheswar}, G., {et~al.} 2012, \mnras, 419, 2587

\bibitem[{{Eswaraiah} {et~al.}(2011){Eswaraiah}, {Pandey}, {Maheswar}, {Medhi}, {Pandey}, {Ojha}, \& {Chen}}]{2011MNRAS.411.1418E}
{Eswaraiah}, C., {Pandey}, A.~K., {Maheswar}, G., {et~al.} 2011, \mnras, 411, 1418

\bibitem[{{Franco} \& {Alves}(2015)}]{2015ApJ...807....5F}
{Franco}, G.~A.~P. \& {Alves}, F.~O. 2015, \apj, 807, 5

\bibitem[{{Franco} {et~al.}(2010){Franco}, {Alves}, \& {Girart}}]{2010ApJ...723..146F}
{Franco}, G.~A.~P., {Alves}, F.~O., \& {Girart}, J.~M. 2010, \apj, 723, 146

\bibitem[{{Gaia Collaboration} {et~al.}(2021){Gaia Collaboration}, {Brown}, {Vallenari}, {Prusti}, {de Bruijne}, {Babusiaux}, {Biermann}, {Creevey}, {Evans}, {Eyer}, {Hutton}, {Jansen}, {Jordi}, {Klioner}, {Lammers}, {Lindegren}, {Luri}, {Mignard}, {Panem}, {Pourbaix}, {Randich}, {Sartoretti}, {Soubiran}, {Walton}, {Arenou}, {Bailer-Jones}, {Bastian}, {Cropper}, {Drimmel}, {Katz}, {Lattanzi}, {van Leeuwen}, {Bakker}, {Cacciari}, {Casta{\~n}eda}, {De Angeli}, {Ducourant}, {Fabricius}, {Fouesneau}, {Fr{\'e}mat}, {Guerra}, {Guerrier}, {Guiraud}, {Jean-Antoine Piccolo}, {Masana}, {Messineo}, {Mowlavi}, {Nicolas}, {Nienartowicz}, {Pailler}, {Panuzzo}, {Riclet}, {Roux}, {Seabroke}, {Sordo}, {Tanga}, {Th{\'e}venin}, {Gracia-Abril}, {Portell}, {Teyssier}, {Altmann}, {Andrae}, {Bellas-Velidis}, {Benson}, {Berthier}, {Blomme}, {Brugaletta}, {Burgess}, {Busso}, {Carry}, {Cellino}, {Cheek}, {Clementini}, {Damerdji}, {Davidson}, {Delchambre}, {Dell'Oro}, {Fern{\'a}ndez-Hern{\'a}ndez}, {Galluccio}, {Garc{\'\i}a-Lario},
  {Garcia-Reinaldos}, {Gonz{\'a}lez-N{\'u}{\~n}ez}, {Gosset}, {Haigron}, {Halbwachs}, {Hambly}, {Harrison}, {Hatzidimitriou}, {Heiter}, {Hern{\'a}ndez}, {Hestroffer}, {Hodgkin}, {Holl}, {Jan{\ss}en}, {Jevardat de Fombelle}, {Jordan}, {Krone-Martins}, {Lanzafame}, {L{\"o}ffler}, {Lorca}, {Manteiga}, {Marchal}, {Marrese}, {Moitinho}, {Mora}, {Muinonen}, {Osborne}, {Pancino}, {Pauwels}, {Petit}, {Recio-Blanco}, {Richards}, {Riello}, {Rimoldini}, {Robin}, {Roegiers}, {Rybizki}, {Sarro}, {Siopis}, {Smith}, {Sozzetti}, {Ulla}, {Utrilla}, {van Leeuwen}, {van Reeven}, {Abbas}, {Abreu Aramburu}, {Accart}, {Aerts}, {Aguado}, {Ajaj}, {Altavilla}, {{\'A}lvarez}, {{\'A}lvarez Cid-Fuentes}, {Alves}, {Anderson}, {Anglada Varela}, {Antoja}, {Audard}, {Baines}, {Baker}, {Balaguer-N{\'u}{\~n}ez}, {Balbinot}, {Balog}, {Barache}, {Barbato}, {Barros}, {Barstow}, {Bartolom{\'e}}, {Bassilana}, {Bauchet}, {Baudesson-Stella}, {Becciani}, {Bellazzini}, {Bernet}, {Bertone}, {Bianchi}, {Blanco-Cuaresma}, {Boch}, {Bombrun}, {Bossini},
  {Bouquillon}, {Bragaglia}, {Bramante}, {Breedt}, {Bressan}, {Brouillet}, {Bucciarelli}, {Burlacu}, {Busonero}, {Butkevich}, {Buzzi}, {Caffau}, {Cancelliere}, {C{\'a}novas}, {Cantat-Gaudin}, {Carballo}, {Carlucci}, {Carnerero}, {Carrasco}, {Casamiquela}, {Castellani}, {Castro-Ginard}, {Castro Sampol}, {Chaoul}, {Charlot}, {Chemin}, {Chiavassa}, {Cioni}, {Comoretto}, {Cooper}, {Cornez}, {Cowell}, {Crifo}, {Crosta}, {Crowley}, {Dafonte}, {Dapergolas}, {David}, {David}, {de Laverny}, {De Luise}, {De March}, {De Ridder}, {de Souza}, {de Teodoro}, {de Torres}, {del Peloso}, {del Pozo}, {Delbo}, {Delgado}, {Delgado}, {Delisle}, {Di Matteo}, {Diakite}, {Diener}, {Distefano}, {Dolding}, {Eappachen}, {Edvardsson}, {Enke}, {Esquej}, {Fabre}, {Fabrizio}, {Faigler}, {Fedorets}, {Fernique}, {Fienga}, {Figueras}, {Fouron}, {Fragkoudi}, {Fraile}, {Franke}, {Gai}, {Garabato}, {Garcia-Gutierrez}, {Garc{\'\i}a-Torres}, {Garofalo}, {Gavras}, {Gerlach}, {Geyer}, {Giacobbe}, {Gilmore}, {Girona}, {Giuffrida}, {Gomel}, {Gomez},
  {Gonzalez-Santamaria}, {Gonz{\'a}lez-Vidal}, {Granvik}, {Guti{\'e}rrez-S{\'a}nchez}, {Guy}, {Hauser}, {Haywood}, {Helmi}, {Hidalgo}, {Hilger}, {H{\l}adczuk}, {Hobbs}, {Holland}, {Huckle}, {Jasniewicz}, {Jonker}, {Juaristi Campillo}, {Julbe}, {Karbevska}, {Kervella}, {Khanna}, {Kochoska}, {Kontizas}, {Kordopatis}, {Korn}, {Kostrzewa-Rutkowska}, {Kruszy{\'n}ska}, {Lambert}, {Lanza}, {Lasne}, {Le Campion}, {Le Fustec}, {Lebreton}, {Lebzelter}, {Leccia}, {Leclerc}, {Lecoeur-Taibi}, {Liao}, {Licata}, {Lindstr{\o}m}, {Lister}, {Livanou}, {Lobel}, {Madrero Pardo}, {Managau}, {Mann}, {Marchant}, {Marconi}, {Marcos Santos}, {Marinoni}, {Marocco}, {Marshall}, {Martin Polo}, {Mart{\'\i}n-Fleitas}, {Masip}, {Massari}, {Mastrobuono-Battisti}, {Mazeh}, {McMillan}, {Messina}, {Michalik}, {Millar}, {Mints}, {Molina}, {Molinaro}, {Moln{\'a}r}, {Montegriffo}, {Mor}, {Morbidelli}, {Morel}, {Morris}, {Mulone}, {Munoz}, {Muraveva}, {Murphy}, {Musella}, {Noval}, {Ord{\'e}novic}, {Orr{\`u}}, {Osinde}, {Pagani}, {Pagano},
  {Palaversa}, {Palicio}, {Panahi}, {Pawlak}, {Pe{\~n}alosa Esteller}, {Penttil{\"a}}, {Piersimoni}, {Pineau}, {Plachy}, {Plum}, {Poggio}, {Poretti}, {Poujoulet}, {Pr{\v{s}}a}, {Pulone}, {Racero}, {Ragaini}, {Rainer}, {Raiteri}, {Rambaux}, {Ramos}, {Ramos-Lerate}, {Re Fiorentin}, {Regibo}, {Reyl{\'e}}, {Ripepi}, {Riva}, {Rixon}, {Robichon}, {Robin}, {Roelens}, {Rohrbasser}, {Romero-G{\'o}mez}, {Rowell}, {Royer}, {Rybicki}, {Sadowski}, {Sagrist{\`a} Sell{\'e}s}, {Sahlmann}, {Salgado}, {Salguero}, {Samaras}, {Sanchez Gimenez}, {Sanna}, {Santove{\~n}a}, {Sarasso}, {Schultheis}, {Sciacca}, {Segol}, {Segovia}, {S{\'e}gransan}, {Semeux}, {Shahaf}, {Siddiqui}, {Siebert}, {Siltala}, {Slezak}, {Smart}, {Solano}, {Solitro}, {Souami}, {Souchay}, {Spagna}, {Spoto}, {Steele}, {Steidelm{\"u}ller}, {Stephenson}, {S{\"u}veges}, {Szabados}, {Szegedi-Elek}, {Taris}, {Tauran}, {Taylor}, {Teixeira}, {Thuillot}, {Tonello}, {Torra}, {Torra}, {Turon}, {Unger}, {Vaillant}, {van Dillen}, {Vanel}, {Vecchiato}, {Viala}, {Vicente},
  {Voutsinas}, {Weiler}, {Wevers}, {Wyrzykowski}, {Yoldas}, {Yvard}, {Zhao}, {Zorec}, {Zucker}, {Zurbach}, \& {Zwitter}}]{2021A&A...649A...1G}
{Gaia Collaboration}, {Brown}, A.~G.~A., {Vallenari}, A., {et~al.} 2021, \aap, 649, A1

\bibitem[{{Garcia-Segura} \& {Franco}(1996)}]{1996ApJ...469..171G}
{Garcia-Segura}, G. \& {Franco}, J. 1996, \apj, 469, 171

\bibitem[{{Hartmann} {et~al.}(2001){Hartmann}, {Ballesteros-Paredes}, \& {Bergin}}]{2001ApJ...562..852H}
{Hartmann}, L., {Ballesteros-Paredes}, J., \& {Bergin}, E.~A. 2001, \apj, 562, 852

\bibitem[{{Hayashi} {et~al.}(2012){Hayashi}, {Itoh}, \& {Oasa}}]{2012PASJ...64...96H}
{Hayashi}, M., {Itoh}, Y., \& {Oasa}, Y. 2012, \pasj, 64, 96

\bibitem[{{Heiles}(2000)}]{2000AJ....119..923H}
{Heiles}, C. 2000, \aj, 119, 923

\bibitem[{{Heitsch} {et~al.}(2001){Heitsch}, {Mac Low}, \& {Klessen}}]{2001ApJ...547..280H}
{Heitsch}, F., {Mac Low}, M.-M., \& {Klessen}, R.~S. 2001, \apj, 547, 280

\bibitem[{{Henney} {et~al.}(2009){Henney}, {Arthur}, {de Colle}, \& {Mellema}}]{2009MNRAS.398..157H}
{Henney}, W.~J., {Arthur}, S.~J., {de Colle}, F., \& {Mellema}, G. 2009, \mnras, 398, 157

\bibitem[{{Hildebrand} {et~al.}(2009){Hildebrand}, {Kirby}, {Dotson}, {Houde}, \& {Vaillancourt}}]{2009ApJ...696..567H}
{Hildebrand}, R.~H., {Kirby}, L., {Dotson}, J.~L., {Houde}, M., \& {Vaillancourt}, J.~E. 2009, \apj, 696, 567

\bibitem[{{H{\o}g} {et~al.}(2000){H{\o}g}, {Fabricius}, {Makarov}, {Urban}, {Corbin}, {Wycoff}, {Bastian}, {Schwekendiek}, \& {Wicenec}}]{2000A&A...355L..27H}
{H{\o}g}, E., {Fabricius}, C., {Makarov}, V.~V., {et~al.} 2000, \aap, 355, L27

\bibitem[{{Hosoya} {et~al.}(2019){Hosoya}, {Itoh}, {Oasa}, {Gupta}, \& {Sen}}]{2019IJAA....9..154H}
{Hosoya}, K., {Itoh}, Y., {Oasa}, Y., {Gupta}, R., \& {Sen}, A.~K. 2019, International Journal of Astronomy and Astrophysics, 9, 154

\bibitem[{{Houde} {et~al.}(2009){Houde}, {Vaillancourt}, {Hildebrand}, {Chitsazzadeh}, \& {Kirby}}]{2009ApJ...706.1504H}
{Houde}, M., {Vaillancourt}, J.~E., {Hildebrand}, R.~H., {Chitsazzadeh}, S., \& {Kirby}, L. 2009, \apj, 706, 1504

\bibitem[{{Juvela} \& {Montillaud}(2016)}]{2016A&A...585A..38J}
{Juvela}, M. \& {Montillaud}, J. 2016, \aap, 585, A38

\bibitem[{{Koenig} {et~al.}(2015){Koenig}, {Hillenbrand}, {Padgett}, \& {DeFelippis}}]{2015AJ....150..100K}
{Koenig}, X., {Hillenbrand}, L.~A., {Padgett}, D.~L., \& {DeFelippis}, D. 2015, \aj, 150, 100

\bibitem[{{Kounkel} {et~al.}(2018){Kounkel}, {Covey}, {Su{\'a}rez}, {Rom{\'a}n-Z{\'u}{\~n}iga}, {Hernandez}, {Stassun}, {Jaehnig}, {Feigelson}, {Pe{\~n}a Ram{\'\i}rez}, {Roman-Lopes}, {Da Rio}, {Stringfellow}, {Kim}, {Borissova}, {Fern{\'a}ndez-Trincado}, {Burgasser}, {Garc{\'\i}a-Hern{\'a}ndez}, {Zamora}, {Pan}, \& {Nitschelm}}]{2018AJ....156...84K}
{Kounkel}, M., {Covey}, K., {Su{\'a}rez}, G., {et~al.} 2018, \aj, 156, 84

\bibitem[{{Lang} {et~al.}(2000){Lang}, {Masheder}, {Dame}, \& {Thaddeus}}]{2000A&A...357.1001L}
{Lang}, W.~J., {Masheder}, M.~R.~W., {Dame}, T.~M., \& {Thaddeus}, P. 2000, \aap, 357, 1001

\bibitem[{Lindegren(2018)}]{LL:LL-124}
Lindegren, L. 2018, gAIA-C3-TN-LU-LL-124

\bibitem[{{Lindegren} {et~al.}(2021){Lindegren}, {Klioner}, {Hern{\'a}ndez}, {Bombrun}, {Ramos-Lerate}, {Steidelm{\"u}ller}, {Bastian}, {Biermann}, {de Torres}, {Gerlach}, {Geyer}, {Hilger}, {Hobbs}, {Lammers}, {McMillan}, {Stephenson}, {Casta{\~n}eda}, {Davidson}, {Fabricius}, {Gracia-Abril}, {Portell}, {Rowell}, {Teyssier}, {Torra}, {Bartolom{\'e}}, {Clotet}, {Garralda}, {Gonz{\'a}lez-Vidal}, {Torra}, {Abbas}, {Altmann}, {Anglada Varela}, {Balaguer-N{\'u}{\~n}ez}, {Balog}, {Barache}, {Becciani}, {Bernet}, {Bertone}, {Bianchi}, {Bouquillon}, {Brown}, {Bucciarelli}, {Busonero}, {Butkevich}, {Buzzi}, {Cancelliere}, {Carlucci}, {Charlot}, {Cioni}, {Crosta}, {Crowley}, {del Peloso}, {del Pozo}, {Drimmel}, {Esquej}, {Fienga}, {Fraile}, {Gai}, {Garcia-Reinaldos}, {Guerra}, {Hambly}, {Hauser}, {Jan{\ss}en}, {Jordan}, {Kostrzewa-Rutkowska}, {Lattanzi}, {Liao}, {Licata}, {Lister}, {L{\"o}ffler}, {Marchant}, {Masip}, {Mignard}, {Mints}, {Molina}, {Mora}, {Morbidelli}, {Murphy}, {Pagani}, {Panuzzo}, {Pe{\~n}alosa
  Esteller}, {Poggio}, {Re Fiorentin}, {Riva}, {Sagrist{\`a} Sell{\'e}s}, {Sanchez Gimenez}, {Sarasso}, {Sciacca}, {Siddiqui}, {Smart}, {Souami}, {Spagna}, {Steele}, {Taris}, {Utrilla}, {van Reeven}, \& {Vecchiato}}]{2021A&A...649A...2L}
{Lindegren}, L., {Klioner}, S.~A., {Hern{\'a}ndez}, J., {et~al.} 2021, \aap, 649, A2

\bibitem[{Liu {et~al.}(2019)Liu, Qiu, Berry, Francesco, Bastien, Koch, Furuya, Kim, Coud√©, Lee, Soam, Eswaraiah, Li, Hwang, Lyo, Pattle, Hasegawa, Kwon, Lai, Ward-Thompson, Ching, Chen, Gu, Li, bai Li, Liu, Qian, Wang, Yuan, Zhang, Zhang, Zhang, Zhou, Zhu, Andr√©, Arzoumanian, Aso, Byun, Chen, Chen, Chen, Cho, Choi, Chrysostomou, Chung, Doi, Drabek-Maunder, Dowell, Eyres, Falle, Fanciullo, Fiege, Franzmann, Friberg, Friesen, Fuller, Gledhill, Graves, Greaves, Griffin, Han, Hatchell, Hayashi, Hoang, Holland, Houde, Inoue, ichiro Inutsuka, Iwasaki, Jeong, Johnstone, Kanamori, hyun Kang, Kang, ju~Kang, Kataoka, Kawabata, Kemper, Kim, Kim, Kim, Kim, Kim, Kirk, Kobayashi, Kusune, Kwon, Lacaille, Lee, Lee, Lee, Lee, Liu, Liu, van Loo, Mairs, Matsumura, Matthews, Moriarty-Schieven, Nagata, Nakamura, Nakanishi, Ohashi, Onaka, Parker, Parsons, Pascale, Peretto, Pon, Pyo, Rao, Rawlings, Retter, Richer, Rigby, Robitaille, Sadavoy, Saito, Savini, Scaife, Seta, Shinnaga, Tamura, Tang, Tomisaka, Tsukamoto, Wang,
  Whitworth, Yen, Yoo, \& Zenko}]{Liu_2019}
Liu, J., Qiu, K., Berry, D., {et~al.} 2019, The Astrophysical Journal, 877, 43

\bibitem[{{Mackey} \& {Lim}(2010)}]{2010MNRAS.403..714M}
{Mackey}, J. \& {Lim}, A.~J. 2010, \mnras, 403, 714

\bibitem[{{Mackey} \& {Lim}(2011)}]{2011MNRAS.412.2079M}
{Mackey}, J. \& {Lim}, A.~J. 2011, \mnras, 412, 2079

\bibitem[{{Magakian} {et~al.}(2004){Magakian}, {Movsessian}, \& {Nikogossian}}]{2004Ap.....47..162M}
{Magakian}, T.~Y., {Movsessian}, T.~A., \& {Nikogossian}, E.~H. 2004, Astrophysics, 47, 162

\bibitem[{{Maheswar} {et~al.}(2010){Maheswar}, {Lee}, {Bhatt}, {Mallik}, \& {Dib}}]{2010A&A...509A..44M}
{Maheswar}, G., {Lee}, C.~W., {Bhatt}, H.~C., {Mallik}, S.~V., \& {Dib}, S. 2010, \aap, 509, A44

\bibitem[{{Mathieu}(2008)}]{2008hsf1.book..757M}
{Mathieu}, R.~D. 2008, {The {$\lambda$} Orionis Star Forming Region}, ed. B.~{Reipurth}, 757

\bibitem[{{McKee} \& {Ostriker}(2007)}]{McKeeOstriker2007}
{McKee}, C.~F. \& {Ostriker}, E.~C. 2007, \araa, 45, 565

\bibitem[{{Miao} {et~al.}(2006){Miao}, {White}, {Nelson}, {Thompson}, \& {Morgan}}]{2006MNRAS.369..143M}
{Miao}, J., {White}, G.~J., {Nelson}, R., {Thompson}, M., \& {Morgan}, L. 2006, \mnras, 369, 143

\bibitem[{{Mouschovias}(1978)}]{Mouschovias1978}
{Mouschovias}, T.~C. 1978, in IAU Colloq. 52: Protostars and Planets, ed. {T.~Gehrels}, 209--242

\bibitem[{{Murdin} \& {Penston}(1977)}]{1977MNRAS.181..657M}
{Murdin}, P. \& {Penston}, M.~V. 1977, \mnras, 181, 657

\bibitem[{{Neha} {et~al.}(2018){Neha}, {Maheswar}, {Soam}, \& {Lee}}]{2018MNRAS.476.4442N}
{Neha}, S., {Maheswar}, G., {Soam}, A., \& {Lee}, C.~W. 2018, \mnras, 476, 4442

\bibitem[{{Neha} {et~al.}(2016){Neha}, {Maheswar}, {Soam}, {Lee}, \& {Tej}}]{2016A&A...588A..45N}
{Neha}, S., {Maheswar}, G., {Soam}, A., {Lee}, C.~W., \& {Tej}, A. 2016, \aap, 588, A45

\bibitem[{{Orsatti} {et~al.}(2010){Orsatti}, {Feinstein}, {Vergne}, {Mart{\'\i}nez}, \& {Vega}}]{2010A&A...513A..75O}
{Orsatti}, A.~M., {Feinstein}, C., {Vergne}, M.~M., {Mart{\'\i}nez}, R.~E., \& {Vega}, E.~I. 2010, \aap, 513, A75

\bibitem[{{Orsatti} {et~al.}(1998){Orsatti}, {Vega}, \& {Marraco}}]{OrsattiVegaMarraco1998}
{Orsatti}, A.~M., {Vega}, E., \& {Marraco}, H.~G. 1998, \aj, 116, 266

\bibitem[{{Ostriker} {et~al.}(2001){Ostriker}, {Stone}, \& {Gammie}}]{2001ApJ...546..980O}
{Ostriker}, E.~C., {Stone}, J.~M., \& {Gammie}, C.~F. 2001, \apj, 546, 980

\bibitem[{{Passot} {et~al.}(1995){Passot}, {Vazquez-Semadeni}, \& {Pouquet}}]{1995ApJ...455..536P}
{Passot}, T., {Vazquez-Semadeni}, E., \& {Pouquet}, A. 1995, \apj, 455, 536

\bibitem[{Pattle {et~al.}(2021)Pattle, Lai, Francesco, Sadavoy, Ward-Thompson, Johnstone, Hoang, Arzoumanian, Bastien, Bourke, Coud√©, Doi, Eswaraiah, Fanciullo, Furuya, Hwang, Hull, Kang, Kim, Kirchschlager, Kwon, Kwon, Lee, Liu, Redman, Soam, Tahani, Tamura, \& Tang}]{Pattle_2021}
Pattle, K., Lai, S.-P., Francesco, J.~D., {et~al.} 2021, The Astrophysical Journal, 907, 88

\bibitem[{{Perryman} {et~al.}(1997){Perryman}, {Lindegren}, {Kovalevsky}, {Hoeg}, {Bastian}, {Bernacca}, {Cr{\'e}z{\'e}}, {Donati}, {Grenon}, {Grewing}, {van Leeuwen}, {van der Marel}, {Mignard}, {Murray}, {Le Poole}, {Schrijver}, {Turon}, {Arenou}, {Froeschl{\'e}}, \& {Petersen}}]{1997A&A...323L..49P}
{Perryman}, M.~A.~C., {Lindegren}, L., {Kovalevsky}, J., {et~al.} 1997, \aap, 323, L49

\bibitem[{{Planck Collaboration I}(2016)}]{2016A&A...594A...1P}
{Planck Collaboration I}. 2016, \aap, 594, A1

\bibitem[{{Ramaprakash} {et~al.}(1998){Ramaprakash}, {Gupta}, {Sen}, \& {Tandon}}]{1998A&AS..128..369R}
{Ramaprakash}, A.~N., {Gupta}, R., {Sen}, A.~K., \& {Tandon}, S.~N. 1998, \aaps, 128, 369

\bibitem[{{Rautela} {et~al.}(2004){Rautela}, {Joshi}, \& {Pandey}}]{2004BASI...32..159R}
{Rautela}, B.~S., {Joshi}, G.~C., \& {Pandey}, J.~C. 2004, Bulletin of the Astronomical Society of India, 32, 159

\bibitem[{{Reipurth}(2000)}]{2000yCat.5104....0R}
{Reipurth}, B. 2000, VizieR Online Data Catalog, 5104

\bibitem[{{Reipurth} \& {Friberg}(2021)}]{2021MNRAS.501.5938R}
{Reipurth}, B. \& {Friberg}, P. 2021, \mnras, 501, 5938

\bibitem[{{Saha} {et~al.}(2022){Saha}, {Gopinathan}, {Ojha}, \& {Neha}}]{2022MNRAS.510.2644S}
{Saha}, P., {Gopinathan}, M., {Ojha}, D.~K., \& {Neha}, S. 2022, \mnras, 510, 2644

\bibitem[{{Santos} {et~al.}(2012){Santos}, {Roman-Lopes}, \& {Franco}}]{2012ApJ...751..138S}
{Santos}, F.~P., {Roman-Lopes}, A., \& {Franco}, G.~A.~P. 2012, \apj, 751, 138

\bibitem[{{Schmidt} {et~al.}(1992){Schmidt}, {Elston}, \& {Lupie}}]{1992AJ....104.1563S}
{Schmidt}, G.~D., {Elston}, R., \& {Lupie}, O.~L. 1992, \aj, 104, 1563

\bibitem[{{Serkowski} {et~al.}(1975){Serkowski}, {Mathewson}, \& {Ford}}]{1975ApJ...196..261S}
{Serkowski}, K., {Mathewson}, D.~S., \& {Ford}, V.~L. 1975, \apj, 196, 261

\bibitem[{{Sharpless}(1959)}]{1959ApJS....4..257S}
{Sharpless}, S. 1959, \apjs, 4, 257

\bibitem[{{Soam} {et~al.}(2013){Soam}, {Maheswar}, {Bhatt}, {Lee}, \& {Ramaprakash}}]{2013MNRAS.432.1502S}
{Soam}, A., {Maheswar}, G., {Bhatt}, H.~C., {Lee}, C.~W., \& {Ramaprakash}, A.~N. 2013, \mnras, 432, 1502

\bibitem[{{Soam} {et~al.}(2015){Soam}, {Maheswar}, {Lee}, {Dib}, {Bhatt}, {Tamura}, \& {Kim}}]{2015A&A...573A..34S}
{Soam}, A., {Maheswar}, G., {Lee}, C.~W., {et~al.} 2015, \aap, 573, A34

\bibitem[{{Soam} {et~al.}(2017){Soam}, {Maheswar}, {Lee}, {Neha}, \& {Andersson}}]{2017MNRAS.465..559S}
{Soam}, A., {Maheswar}, G., {Lee}, C.~W., {Neha}, S., \& {Andersson}, B.~G. 2017, \mnras, 465, 559

\bibitem[{{Soam} {et~al.}(2018){Soam}, {Maheswar}, {Lee}, {Neha}, \& {Kim}}]{2018MNRAS.476.4782S}
{Soam}, A., {Maheswar}, G., {Lee}, C.~W., {Neha}, S., \& {Kim}, K.-T. 2018, \mnras, 476, 4782

\bibitem[{{Soler} {et~al.}(2013){Soler}, {Hennebelle}, {Martin}, {Miville-Desch{\^e}nes}, {Netterfield}, \& {Fissel}}]{2013ApJ...774..128S}
{Soler}, J.~D., {Hennebelle}, P., {Martin}, P.~G., {et~al.} 2013, \apj, 774, 128

\bibitem[{{Sugitani} {et~al.}(1991){Sugitani}, {Fukui}, \& {Ogura}}]{1991ApJS...77...59S}
{Sugitani}, K., {Fukui}, Y., \& {Ogura}, K. 1991, \apjs, 77, 59

\bibitem[{{Topasna} {et~al.}(2017){Topasna}, {Daman}, \& {Kaltcheva}}]{2017PASP..129j4201T}
{Topasna}, G.~A., {Daman}, E.~A., \& {Kaltcheva}, N.~T. 2017, \pasp, 129, 104201

\bibitem[{{van Leeuwen}(2007)}]{2007A&A...474..653V}
{van Leeuwen}, F. 2007, \aap, 474, 653

\bibitem[{{Vergne} {et~al.}(2007){Vergne}, {Feinstein}, \& {Mart{\'{\i}}nez}}]{2007A&A...462..621V}
{Vergne}, M.~M., {Feinstein}, C., \& {Mart{\'{\i}}nez}, R. 2007, \aap, 462, 621

\bibitem[{{Wade}(1957)}]{1957AJ.....62..148W}
{Wade}, C.~M. 1957, \aj, 62, 148

\bibitem[{{Whittet}(1977)}]{Whittet1977}
{Whittet}, D.~C.~B. 1977, \mnras, 180, 29

\bibitem[{{Whittet} \& {van Breda}(1978)}]{1978A&A....66...57W}
{Whittet}, D.~C.~B. \& {van Breda}, I.~G. 1978, \aap, 66, 57

\bibitem[{{Wu} {et~al.}(2004){Wu}, {Wei}, {Zhao}, {Shi}, {Yu}, {Qin}, \& {Huang}}]{2004A&A...426..503W}
{Wu}, Y., {Wei}, Y., {Zhao}, M., {et~al.} 2004, \aap, 426, 503

\bibitem[{{Zamora-Avil{\'e}s} {et~al.}(2019){Zamora-Avil{\'e}s}, {V{\'a}zquez-Semadeni}, {Gonz{\'a}lez}, {Franco}, {Shore}, {Hartmann}, {Ballesteros-Paredes}, {Banerjee}, \& {K{\"o}rtgen}}]{2019MNRAS.487.2200Z}
{Zamora-Avil{\'e}s}, M., {V{\'a}zquez-Semadeni}, E., {Gonz{\'a}lez}, R.~F., {et~al.} 2019, \mnras, 487, 2200

\bibitem[{{Zari} {et~al.}(2018){Zari}, {Hashemi}, {Brown}, {Jardine}, \& {de Zeeuw}}]{2018A&A...620A.172Z}
{Zari}, E., {Hashemi}, H., {Brown}, A.~G.~A., {Jardine}, K., \& {de Zeeuw}, P.~T. 2018, \aap, 620, A172

\bibitem[{{Zhang} {et~al.}(1989){Zhang}, {Laureijs}, {Chlewicki}, {Wesselius}, \& {Clark}}]{1989A&A...218..231Z}
{Zhang}, C.~Y., {Laureijs}, R.~J., {Chlewicki}, G., {Wesselius}, P.~R., \& {Clark}, F.~O. 1989, \aap, 218, 231

\bibitem[{{Zhou} {et~al.}(1988){Zhou}, {Butner}, \& {Evans}}]{1988ApJ...333..809Z}
{Zhou}, S., {Butner}, H.~M., \& {Evans}, Neal~J., I. 1988, \apj, 333, 809

\end{thebibliography}

\begin{appendix} 
\section{Additional table}
\begin{table*}
\caption{Polarization results of observed Tycho stars in VRI filters towards $\lambda$-Ori region.}\label{tab:Tychos}
\begin{center}
\resizebox{1.0\linewidth}{!}{%
\begin{tabular}{lcrcrcrcrcccr}\hline
ID &    RA  	  &  Dec  	& P$_{V}$ $\pm$ $\epsilon_{P_{V}}$ & $\theta_{V}$ $\pm$ $\epsilon_{\theta_{V}}$  &P$_{R}$ $\pm$ $\epsilon_{P_{R}}$ & $\theta_{R}$ $\pm$ $\epsilon_{\theta_{R}}$  & P$_{I}$ $\pm$ $\epsilon_{P_{I}}$ & $\theta_{I}$ $\pm$ $\epsilon_{\theta_{I}}$ & $\lambda_{max}$ $\pm$ $\epsilon_{\lambda_{max}}$ & P$_{max}$ $\pm$ $\epsilon_{P_{max}}$ & R$_{V}$ $\pm$ $\epsilon_{R_{V}}$ & D $\pm$ $\epsilon_D$ \\  
  &($\degree$)&($\degree$)& (\%) 			&($\degree$)  & (\%) 			&($\degree$)  & (\%) 			&($\degree$)  & ($\mu$m)  & (\%) & & (pc) \\
 (1) & (2) & (3) & (4) & (5) & (6) & (7) & (8) & (9) & (10) & (11) & (12) & (13) \\ \hline
1 & 79.683675 & 11.227991 & 0.5 $\pm$ 0.2 & 101 $\pm$ 6 & 0.4 $\pm$ 0.1 & 134 $\pm$ 7 & 0.4 $\pm$ 0.2 & 133 $\pm$ 7 & 0.56 $\pm$ 0.06 & 0.46 $\pm$ 0.02 & 3.14 $\pm$ 0.38 & 1059 $\pm$ 19 \\ 
2 & 80.334997 & 6.959988 & 1.4 $\pm$ 0.2 & 123 $\pm$ 5 & 1.3 $\pm$ 0.1 & 126 $\pm$ 6 & 1.2 $\pm$ 0.1 & 127 $\pm$ 6 & 0.60 $\pm$ 0.04 & 1.35 $\pm$ 0.04 & 3.36 $\pm$ 0.27 & 1601 $\pm$ 67 \\ 
3 & 80.450959 & 13.120884 & 0.8 $\pm$ 0.2 & 111 $\pm$ 5 & 0.8 $\pm$ 0.2 & 119 $\pm$ 6 & 0.6 $\pm$ 0.2 & 115 $\pm$ 6 & 0.48 $\pm$ 0.03 & 0.85 $\pm$ 0.05 & 2.69 $\pm$ 0.24 & 939 $\pm$ 25 \\ 
4 & 81.025895 & 11.663095 & 0.5 $\pm$ 0.2 & 82 $\pm$ 7 & 0.3 $\pm$ 0.2 & 95 $\pm$ 7 & 0.3 $\pm$ 0.2 & 100 $\pm$ 7 & 0.41 $\pm$ 0.07 & 0.48 $\pm$ 0.08 & 2.30 $\pm$ 0.40 & 584 $\pm$ 7 \\ 
5 & 81.100961 & 7.859315 & 0.5 $\pm$ 0.2 & 74 $\pm$ 5 & 0.5 $\pm$ 0.2 & 96 $\pm$ 7 & 0.3 $\pm$ 0.2 & 92 $\pm$ 8 & 0.39 $\pm$ 0.07 & 0.59 $\pm$ 0.12 & 2.18 $\pm$ 0.38 & 551 $\pm$ 8 \\ 
6 & 81.380420 & 9.098704 & 0.7 $\pm$ 0.2 & 170 $\pm$ 7 & 0.5 $\pm$ 0.2 & 159 $\pm$ 7 & 0.5 $\pm$ 0.2 & 157 $\pm$ 7 & 0.41 $\pm$ 0.06 & 0.77 $\pm$ 0.12 & 2.30 $\pm$ 0.35 & 696 $\pm$ 8 \\ 
7 & 82.017076 & 11.218365 & 1.1 $\pm$ 0.2 & 147 $\pm$ 5 & 1.1 $\pm$ 0.2 & 149 $\pm$ 5 & 1.0 $\pm$ 0.2 & 46 $\pm$ 6 & 0.58 $\pm$ 0.04 & 1.14 $\pm$ 0.04 & 3.25 $\pm$ 0.29 & 1214 $\pm$ 26 \\ 
8 & 82.475707 & 5.422993 & 1.2 $\pm$ 0.2 & 158 $\pm$ 5 & 1.1 $\pm$ 0.1 & 157 $\pm$ 7 & 0.9 $\pm$ 0.1 & 149 $\pm$ 6 & 0.51 $\pm$ 0.03 & 1.17 $\pm$ 0.05 & 2.86 $\pm$ 0.21 & 1921 $\pm$ 73 \\ 
9 & 82.609959 & 9.636632 & 0.5 $\pm$ 0.2 & 154 $\pm$ 7 & 0.4 $\pm$ 0.1 & 147 $\pm$ 7 & 0.4 $\pm$ 0.1 & 160 $\pm$ 7 & 0.60 $\pm$ 0.05 & 0.46 $\pm$ 0.02 & 3.36 $\pm$ 0.32 & 946 $\pm$ 16 \\ 
10 & 82.619749 & 8.075273 & 0.8 $\pm$ 0.2 & 149 $\pm$ 5 & 0.7 $\pm$ 0.2 & 135 $\pm$ 6 & 0.6 $\pm$ 0.2 & 130 $\pm$ 6 & 0.47 $\pm$ 0.04 & 0.82 $\pm$ 0.06 & 2.63 $\pm$ 0.27 & 988 $\pm$ 18 \\ 
11 & 82.713199 & 14.039175 & 1.5 $\pm$ 0.2 & 43 $\pm$ 5 & 1.4 $\pm$ 0.2 & 41 $\pm$ 6 & 1.1 $\pm$ 0.2 & 33 $\pm$ 7 & 0.46 $\pm$ 0.02 & 1.56 $\pm$ 0.05 & 2.58 $\pm$ 0.17 & 919 $\pm$ 17 \\ 
12 & 83.158619 & 9.086116 & 1.4 $\pm$ 0.2 & 153 $\pm$ 4 & 1.4 $\pm$ 0.1 & 149 $\pm$ 7 & 1.3 $\pm$ 0.1 & 144 $\pm$ 7 & 0.60 $\pm$ 0.00 & 1.41 $\pm$ 0.00 & 3.36 $\pm$ 0.18 & 1161 $\pm$ 65 \\ 
13 & 83.325927 & 6.805111 & 0.9 $\pm$ 0.2 & 142 $\pm$ 6 & 0.8 $\pm$ 0.2 & 149 $\pm$ 6 & 0.7 $\pm$ 0.2 & 148 $\pm$ 7 & 0.53 $\pm$ 0.01 & 0.88 $\pm$ 0.01 & 2.97 $\pm$ 0.17 & 932 $\pm$ 17 \\ 
14 & 83.361726 & 7.842497 & 0.7 $\pm$ 0.2 & 167 $\pm$ 7 & 0.5 $\pm$ 0.2 & 156 $\pm$ 7 & 0.5 $\pm$ 0.2 & 152 $\pm$ 9 & 0.40 $\pm$ 0.09 & 0.79 $\pm$ 0.18 & 2.24 $\pm$ 0.49 & 1399 $\pm$ 56 \\ 
15 & 83.483675 & 10.600003 & 0.9 $\pm$ 0.2 & 162 $\pm$ 5 & 0.7 $\pm$ 0.1 & 142 $\pm$ 7 & 0.7 $\pm$ 0.2 & 158 $\pm$ 7 & 0.44 $\pm$ 0.14 & 0.92 $\pm$ 0.27 & 2.46 $\pm$ 0.79 & 1267 $\pm$ 30 \\ 
16 & 83.580072 & 11.654027 & 0.7 $\pm$ 0.2 & 176 $\pm$ 7 & 0.6 $\pm$ 0.2 & 165 $\pm$ 7 & 0.5 $\pm$ 0.2 & 175 $\pm$ 9 & 0.45 $\pm$ 0.01 & 0.70 $\pm$ 0.02 & 2.52 $\pm$ 0.16 & 1096 $\pm$ 23 \\ 
17 & 83.713086 & 5.758289 & 0.7 $\pm$ 0.2 & 13 $\pm$ 11 & 0.5 $\pm$ 0.2 & 11 $\pm$ 13 & 0.6 $\pm$ 0.2 & 12 $\pm$ 7 & 0.46 $\pm$ 0.13 & 0.72 $\pm$ 0.17 & 2.58 $\pm$ 0.76 & 1090 $\pm$ 38 \\ 
18 & 84.201449 & 10.201707 & 2.6 $\pm$ 0.1 & 143 $\pm$ 4 & 2.3 $\pm$ 0.1 & 142 $\pm$ 8 & 2.2 $\pm$ 0.1 & 145 $\pm$ 8 & 0.54 $\pm$ 0.04 & 2.53 $\pm$ 0.11 & 3.02 $\pm$ 0.29 & 905 $\pm$ 20 \\ 
19 & 84.329903 & 11.133247 & 1.1 $\pm$ 0.2 & 173 $\pm$ 5 & 0.9 $\pm$ 0.1 & 172 $\pm$ 6 & 0.9 $\pm$ 0.1 & 177 $\pm$ 6 & 0.55 $\pm$ 0.16 & 1.00 $\pm$ 0.19 & 3.08 $\pm$ 0.93 & 1274 $\pm$ 30 \\ 
20 & 84.458219 & 9.303417 & 1.1 $\pm$ 0.2 & 169 $\pm$ 5 & 1.0 $\pm$ 0.1 & 169 $\pm$ 7 & 1.0 $\pm$ 0.1 & 168 $\pm$ 6 & 0.66 $\pm$ 0.06 & 1.05 $\pm$ 0.03 & 3.70 $\pm$ 0.41 & 2220 $\pm$ 91 \\ 
21 & 84.851907 & 14.632194 & 1.8 $\pm$ 0.2 & 171 $\pm$ 5 & 1.6 $\pm$ 0.1 & 167 $\pm$ 7 & 1.6 $\pm$ 0.2 & 168 $\pm$ 6 & 0.55 $\pm$ 0.08 & 1.75 $\pm$ 0.13 & 3.08 $\pm$ 0.50 & 1092 $\pm$ 32 \\ 
22 & 84.882692 & 9.287090 & 0.9 $\pm$ 0.3 & 160 $\pm$ 7 & 0.9 $\pm$ 0.1 & 168 $\pm$ 6 & 0.9 $\pm$ 0.2 & 167 $\pm$ 7 & 0.66 $\pm$ 0.03 & 0.90 $\pm$ 0.01 & 3.70 $\pm$ 0.26 & 1272 $\pm$ 39 \\ 
23 & 84.989390 & 9.698935 & 1.0 $\pm$ 0.1 & 174 $\pm$ 6 & 1.0 $\pm$ 0.1 & 171 $\pm$ 6 & 0.9 $\pm$ 0.1 & 170 $\pm$ 6 & 0.60 $\pm$ 0.01 & 0.97 $\pm$ 0.00 & 3.36 $\pm$ 0.18 & 940 $\pm$ 16 \\ 
24 & 85.054662 & 8.675390 & 1.4 $\pm$ 0.2 & 146 $\pm$ 4 & 1.4 $\pm$ 0.2 & 150 $\pm$ 6 & 1.2 $\pm$ 0.2 & 48 $\pm$ 6 & 0.57 $\pm$ 0.02 & 1.36 $\pm$ 0.03 & 3.19 $\pm$ 0.22 & 1421 $\pm$ 37 \\ 
25 & 85.102108 & 10.114206 & 1.6 $\pm$ 0.2 & 172 $\pm$ 5 & 1.2 $\pm$ 0.1 & 163 $\pm$ 6 & 1.1 $\pm$ 0.1 & 164 $\pm$ 6 & 0.48 $\pm$ 0.13 & 1.45 $\pm$ 0.35 & 2.69 $\pm$ 0.75 & 1957 $\pm$ 89 \\ 
26 & 85.354677 & 14.360165 & 2.1 $\pm$ 0.2 & 170 $\pm$ 5 & 1.8 $\pm$ 0.1 & 170 $\pm$ 7 & 1.9 $\pm$ 0.1 & 169 $\pm$ 6 & 0.63 $\pm$ 0.13 & 1.93 $\pm$ 0.16 & 3.53 $\pm$ 0.75 & 646 $\pm$ 9 \\ 
27 & 85.846172 & 10.807565 & 2.2 $\pm$ 0.2 & 4 $\pm$ 5 & 2.0 $\pm$ 0.2 & 5 $\pm$ 5 & 1.4 $\pm$ 0.2 & 79 $\pm$ 6 & 0.40 $\pm$ 0.04 & 2.59 $\pm$ 0.32 & 2.24 $\pm$ 0.27 & 1194 $\pm$ 24 \\ 
28 & 86.410384 & 13.849815 & 2.9 $\pm$ 0.2 & 156 $\pm$ 4 & 2.6 $\pm$ 0.2 & 154 $\pm$ 6 & 2.4 $\pm$ 0.1 & 154 $\pm$ 6 & 0.55 $\pm$ 0.03 & 2.84 $\pm$ 0.10 & 3.08 $\pm$ 0.23 & 910 $\pm$ 21 \\ 
29 & 86.495363 & 8.237844 & 1.6 $\pm$ 0.2 & 174 $\pm$ 4 & 1.4 $\pm$ 0.2 & 174 $\pm$ 7 & 1.2 $\pm$ 0.2 & 174 $\pm$ 7 & 0.51 $\pm$ 0.00 & 1.56 $\pm$ 0.00 & 2.86 $\pm$ 0.15 & 985 $\pm$ 27 \\ 
30 & 87.914164 & 7.541778 & 0.9 $\pm$ 0.1 & 14 $\pm$ 6 & 0.8 $\pm$ 0.1 & 10 $\pm$ 9 & 0.8 $\pm$ 0.1 & 8 $\pm$ 6 & 0.54 $\pm$ 0.07 & 0.89 $\pm$ 0.06 & 3.02 $\pm$ 0.41 & 807 $\pm$ 14 \\ 
31 & 88.040171 & 7.731424 & 0.8 $\pm$ 0.2 & 168 $\pm$ 11 & 0.7 $\pm$ 0.2 & 172 $\pm$ 7 & 0.4 $\pm$ 0.2 & 3 $\pm$ 7 & 0.35 $\pm$ 0.12 & 1.00 $\pm$ 0.48 & 1.96 $\pm$ 0.69 & 944 $\pm$ 18 \\ 
32 & 88.091171 & 12.577652 & 1.9 $\pm$ 0.1 & 4 $\pm$ 7 & 1.6 $\pm$ 0.1 & 171 $\pm$ 6 & 1.5 $\pm$ 0.1 & 174 $\pm$ 7 & 0.48 $\pm$ 0.08 & 1.92 $\pm$ 0.23 & 2.69 $\pm$ 0.45 & 1035 $\pm$ 27 \\ 
33 & 88.442286 & 7.989944 & 1.0 $\pm$ 0.2 & 35 $\pm$ 8 & 0.8 $\pm$ 0.2 & 27 $\pm$ 9 & 0.6 $\pm$ 0.2 & 23 $\pm$ 10 & 0.38 $\pm$ 0.01 & 1.20 $\pm$ 0.03 & 2.13 $\pm$ 0.12 & 1893 $\pm$ 106 \\ 
34 & 88.475476 & 10.499152 & 3.5 $\pm$ 0.2 & 167 $\pm$ 5 & 3.1 $\pm$ 0.2 & 164 $\pm$ 6 & 3.0 $\pm$ 0.2 & 164 $\pm$ 6 & 0.55 $\pm$ 0.04 & 3.41 $\pm$ 0.12 & 3.08 $\pm$ 0.27 & 1152 $\pm$ 24 \\ 
\hline
\end{tabular} }
\tablefoot{Columns are listed as follows: (1) serial number; (2) RA; (3) Dec; (4), (6) \& (8) degree of polarization with their respective errors in V, R \& I bands, respectively; (5), (7) \& (9) polarization position angles with their respective errors in V, R \& I bands, respectively; (10)$-$(11) maximum wavelength and degree of polarization estimated from Serkowski fitting; (12) total-to-selective extinction; (13) distances with their respective errors from \citealt{Bailer_Jones_2021}.}
\end{center}
\end{table*}


\begin{table*}   
\begin{center}        
\caption{Polarization results of 105 stars observed in the direction of BRC 17 and 171 stars observed towards BRC 18 with P/$\sigma_{P}$ $\ge$ 2 in R filter.}\label{tab:Polresult}
\begin{tabular}{cc}                                               
\begin{tabular}[t]{ccccr}\hline                                   
Star & $\alpha$ (J2000)  & $\delta$ (J2000)  & P $\pm$ $\epsilon_P$ & $\theta$ $\pm$ $\epsilon_{\theta}$  \\ 
 Id  &   ($\degree$) &    ($\degree$)&    (\%)           & ($\degree$) \\\hline    
\multicolumn{5}{c}{\bf BRC 17} \\                                  
1	 & 	82.568436	 & 	12.238192	 & 	0.9 $\pm$ 0.2	 & 	23 $\pm$ 8 \\ 
2	 & 	82.599312	 & 	12.227496	 & 	2.3 $\pm$ 0.9	 & 	14 $\pm$ 13 \\ 
3	 & 	82.601128	 & 	12.094558	 & 	1.3 $\pm$ 0.4	 & 	26 $\pm$ 11 \\ 
4	 & 	82.601219	 & 	12.198106	 & 	0.8 $\pm$ 0.2	 & 	50 $\pm$ 8 \\ 
5	 & 	82.602654	 & 	12.087220	 & 	0.8 $\pm$ 0.1	 & 	155 $\pm$ 5 \\ 
6	 & 	82.603317	 & 	12.230099	 & 	3.3 $\pm$ 0.6	 & 	44 $\pm$ 6 \\ 
7	 & 	82.604851	 & 	12.068765	 & 	1.9 $\pm$ 0.6	 & 	165 $\pm$ 9 \\ 
8	 & 	82.616081	 & 	12.116481	 & 	1.8 $\pm$ 0.6	 & 	22 $\pm$ 10 \\ 
9	 & 	82.627769	 & 	12.216683	 & 	0.7 $\pm$ 0.2	 & 	19 $\pm$ 8 \\ 
10	 & 	82.634521	 & 	12.121438	 & 	1.1 $\pm$ 0.3	 & 	152 $\pm$ 7 \\ 
11	 & 	82.638466	 & 	12.123507	 & 	0.4 $\pm$ 0.2	 & 	164 $\pm$ 8 \\ 
12	 & 	82.646278	 & 	12.116957	 & 	0.6 $\pm$ 0.2	 & 	133 $\pm$ 6 \\ 
13	 & 	82.647423	 & 	12.222106	 & 	1.1 $\pm$ 0.3	 & 	21 $\pm$ 11 \\ 
14	 & 	82.648911	 & 	12.214931	 & 	1.8 $\pm$ 0.6	 & 	60 $\pm$ 10 \\ 
15	 & 	82.660690	 & 	12.138010	 & 	1.7 $\pm$ 0.8	 & 	111 $\pm$ 11 \\ 
16	 & 	82.661232	 & 	12.234212	 & 	0.6 $\pm$ 0.1	 & 	14 $\pm$ 8 \\ 
17	 & 	82.667831	 & 	12.075129	 & 	1.6 $\pm$ 0.6	 & 	127 $\pm$ 9 \\ 
18	 & 	82.676125	 & 	12.123517	 & 	1.7 $\pm$ 0.6	 & 	148 $\pm$ 10 \\ 
19	 & 	82.677361	 & 	12.146648	 & 	1.3 $\pm$ 0.3	 & 	1 $\pm$ 7 \\ 
20	 & 	82.692673	 & 	12.224899	 & 	1.4 $\pm$ 0.5	 & 	29 $\pm$ 12 \\ 
21	 & 	82.695663	 & 	12.190128	 & 	1.4 $\pm$ 0.2	 & 	26 $\pm$ 5 \\ 
22	 & 	82.698982	 & 	12.070744	 & 	2.4 $\pm$ 0.5	 & 	102 $\pm$ 5 \\ 
23	 & 	82.700195	 & 	12.076781	 & 	0.5 $\pm$ 0.1	 & 	7 $\pm$ 10 \\ 
24	 & 	82.700844	 & 	12.086476	 & 	0.8 $\pm$ 0.2	 & 	170 $\pm$ 8 \\ 
25	 & 	82.702469	 & 	12.060691	 & 	0.9 $\pm$ 0.2	 & 	5 $\pm$ 7 \\ 
26	 & 	82.702957	 & 	12.115624	 & 	1.0 $\pm$ 0.2	 & 	99 $\pm$ 6 \\ 
27	 & 	82.708870	 & 	12.030152	 & 	1.1 $\pm$ 0.4	 & 	66 $\pm$ 8 \\ 
28	 & 	82.714439	 & 	12.097614	 & 	0.7 $\pm$ 0.2	 & 	10 $\pm$ 8 \\ 
29	 & 	82.715256	 & 	12.156046	 & 	0.5 $\pm$ 0.2	 & 	113 $\pm$ 7 \\ 
30	 & 	82.724586	 & 	12.007061	 & 	0.5 $\pm$ 0.1	 & 	150 $\pm$ 5 \\ 
31	 & 	82.733559	 & 	12.025238	 & 	1.1 $\pm$ 0.3	 & 	113 $\pm$ 5 \\ 
32	 & 	82.739586	 & 	12.017054	 & 	1.5 $\pm$ 0.7	 & 	94 $\pm$ 11 \\ 
33	 & 	82.743492	 & 	12.009210	 & 	0.4 $\pm$ 0.2	 & 	111 $\pm$ 6 \\ 
34	 & 	82.747086	 & 	12.030783	 & 	0.7 $\pm$ 0.3	 & 	86 $\pm$ 7 \\ 
35	 & 	82.747955	 & 	12.210904	 & 	2.3 $\pm$ 0.9	 & 	28 $\pm$ 13 \\ 
36	 & 	82.750214	 & 	12.070655	 & 	1.6 $\pm$ 0.4	 & 	75 $\pm$ 6 \\ 
37	 & 	82.751984	 & 	12.215071	 & 	1.1 $\pm$ 0.1	 & 	70 $\pm$ 5 \\ 
38	 & 	82.754005	 & 	12.083478	 & 	1.0 $\pm$ 0.3	 & 	72 $\pm$ 5 \\ 
39	 & 	82.754349	 & 	11.992623	 & 	0.7 $\pm$ 0.2	 & 	149 $\pm$ 7 \\ 
40	 & 	82.756210	 & 	11.996511	 & 	1.0 $\pm$ 0.2	 & 	146 $\pm$ 5 \\ 

41	 & 	82.764862	 & 	12.166323	 & 	1.6 $\pm$ 0.2	 & 	19 $\pm$ 6 \\ 
42	 & 	82.766769	 & 	12.180549	 & 	2.2 $\pm$ 0.2	 & 	17 $\pm$ 4 \\ 
43	 & 	82.772522	 & 	11.981793	 & 	0.9 $\pm$ 0.3	 & 	136 $\pm$ 6 \\ 
44	 & 	82.774445	 & 	12.002536	 & 	1.0 $\pm$ 0.5	 & 	118 $\pm$ 10 \\ 
45	 & 	82.776337	 & 	12.075002	 & 	1.0 $\pm$ 0.4	 & 	43 $\pm$ 11 \\ 
46	 & 	82.784729	 & 	12.160297	 & 	1.0 $\pm$ 0.2	 & 	158 $\pm$ 7 \\ 
47	 & 	82.788124	 & 	12.295055	 & 	1.9 $\pm$ 0.6	 & 	47 $\pm$ 8 \\ 
48	 & 	82.793694	 & 	12.319737	 & 	2.1 $\pm$ 0.3	 & 	85 $\pm$ 4 \\ 
49	 & 	82.796494	 & 	12.039086	 & 	0.3 $\pm$ 0.1	 & 	177 $\pm$ 8 \\ 
50	 & 	82.799156	 & 	12.253103	 & 	0.8 $\pm$ 0.2	 & 	178 $\pm$ 7 \\ 
\hline
\end{tabular}
\begin{tabular}[t]{ccccr}  \hline
Star  & $\alpha$ (J2000)  & $\delta$ (J2000)  & P $\pm$ $\epsilon_P$ & $\theta$ $\pm$ $\epsilon_{\theta}$  \\ 
Id  &($\degree$)&($\degree$)& (\%) &($\degree$) \\\hline
\multicolumn{5}{c}{}\\
51	 & 	82.803902	 & 	12.133664	 & 	2.6 $\pm$ 0.1	 & 	19 $\pm$ 4 \\ 
52	 & 	82.808525	 & 	12.271837	 & 	2.0 $\pm$ 1.0	 & 	110 $\pm$ 12 \\
53	 & 	82.812210	 & 	12.318285	 & 	2.8 $\pm$ 0.4	 & 	57 $\pm$ 4 \\ 
54	 & 	82.814453	 & 	12.202498	 & 	2.3 $\pm$ 0.4	 & 	118 $\pm$ 4 \\ 
55	 & 	82.827515	 & 	12.333557	 & 	1.1 $\pm$ 0.4	 & 	55 $\pm$ 8 \\ 
56	 & 	82.828484	 & 	12.260856	 & 	1.6 $\pm$ 0.5	 & 	180 $\pm$ 11 \\
57	 & 	82.830978	 & 	12.166741	 & 	2.5 $\pm$ 0.3	 & 	144 $\pm$ 4 \\ 
58	 & 	82.835464	 & 	12.167185	 & 	0.7 $\pm$ 0.3	 & 	130 $\pm$ 9 \\ 
59	 & 	82.837013	 & 	12.071788	 & 	2.2 $\pm$ 0.8	 & 	77 $\pm$ 9 \\ 
60	 & 	82.838844	 & 	12.283099	 & 	0.8 $\pm$ 0.3	 & 	127 $\pm$ 7 \\ 
61	 & 	82.840187	 & 	12.096471	 & 	0.3 $\pm$ 0.1	 & 	102 $\pm$ 4 \\ 
62	 & 	82.856598	 & 	12.245782	 & 	2.2 $\pm$ 0.1	 & 	179 $\pm$ 5 \\ 
63	 & 	82.859009	 & 	12.292132	 & 	2.1 $\pm$ 0.2	 & 	148 $\pm$ 3 \\ 
64	 & 	82.869324	 & 	12.046968	 & 	1.4 $\pm$ 0.4	 & 	140 $\pm$ 6 \\ 
65	 & 	82.881615	 & 	12.048403	 & 	1.5 $\pm$ 0.6	 & 	44 $\pm$ 12 \\ 
66	 & 	82.906693	 & 	12.028670	 & 	0.4 $\pm$ 0.2	 & 	158 $\pm$ 9 \\ 
67	 & 	82.914528	 & 	12.183977	 & 	2.7 $\pm$ 0.8	 & 	138 $\pm$ 8 \\ 
68	 & 	82.922920	 & 	12.217783	 & 	1.6 $\pm$ 0.5	 & 	150 $\pm$ 8 \\ 
69	 & 	82.924316	 & 	12.098646	 & 	2.1 $\pm$ 0.5	 & 	110 $\pm$ 6 \\ 
70	 & 	82.928040	 & 	12.241245	 & 	3.5 $\pm$ 1.5	 & 	164 $\pm$ 12 \\
71	 & 	82.932304	 & 	12.028439	 & 	2.8 $\pm$ 0.8	 & 	89 $\pm$ 7 \\ 
72	 & 	82.943199	 & 	12.228580	 & 	0.7 $\pm$ 0.2	 & 	175 $\pm$ 7 \\ 
73	 & 	82.948074	 & 	12.319831	 & 	0.7 $\pm$ 0.3	 & 	154 $\pm$ 11 \\
74	 & 	82.949745	 & 	12.182503	 & 	1.2 $\pm$ 0.4	 & 	70 $\pm$ 6 \\ 
75	 & 	82.952950	 & 	12.171328	 & 	1.2 $\pm$ 0.5	 & 	157 $\pm$ 10 \\
76	 & 	82.975975	 & 	12.368179	 & 	1.1 $\pm$ 0.2	 & 	18 $\pm$ 7 \\ 
77	 & 	82.984650	 & 	12.411470	 & 	2.6 $\pm$ 0.2	 & 	7 $\pm$ 3 \\ 
78	 & 	82.984703	 & 	12.239669	 & 	1.8 $\pm$ 0.4	 & 	171 $\pm$ 6 \\ 
79	 & 	82.989380	 & 	12.106259	 & 	0.4 $\pm$ 0.1	 & 	0 $\pm$ 9 \\ 
80	 & 	82.989792	 & 	12.209799	 & 	0.5 $\pm$ 0.1	 & 	53 $\pm$ 6 \\ 
81	 & 	82.991608	 & 	12.101281	 & 	1.8 $\pm$ 0.6	 & 	167 $\pm$ 9 \\ 
82	 & 	82.994530	 & 	12.370210	 & 	2.3 $\pm$ 0.5	 & 	159 $\pm$ 5 \\ 
83	 & 	82.995323	 & 	12.180933	 & 	0.4 $\pm$ 0.1	 & 	169 $\pm$ 9 \\ 
84	 & 	82.996674	 & 	12.244583	 & 	0.8 $\pm$ 0.3	 & 	171 $\pm$ 11 \\
85	 & 	83.002495	 & 	12.126358	 & 	1.4 $\pm$ 0.4	 & 	158 $\pm$ 8 \\ 
86	 & 	83.007202	 & 	12.057828	 & 	0.5 $\pm$ 0.1	 & 	2 $\pm$ 8 \\ 
87	 & 	83.008606	 & 	12.207569	 & 	1.2 $\pm$ 0.6	 & 	105 $\pm$ 11 \\
88	 & 	83.014168	 & 	12.470267	 & 	2.4 $\pm$ 0.4	 & 	27 $\pm$ 5 \\ 
89	 & 	83.018013	 & 	12.199817	 & 	2.3 $\pm$ 0.8	 & 	132 $\pm$ 10 \\
90	 & 	83.019943	 & 	12.425723	 & 	1.4 $\pm$ 0.2	 & 	159 $\pm$ 5 \\ 
91	 & 	83.022797	 & 	12.404117	 & 	3.1 $\pm$ 1.3	 & 	157 $\pm$ 11 \\
92	 & 	83.023544	 & 	12.399603	 & 	2.0 $\pm$ 0.6	 & 	152 $\pm$ 7 \\ 
93	 & 	83.030342	 & 	12.153024	 & 	1.4 $\pm$ 0.5	 & 	108 $\pm$ 8 \\ 
94	 & 	83.034149	 & 	12.339017	 & 	3.4 $\pm$ 0.3	 & 	15 $\pm$ 3 \\ 
95	 & 	83.034348	 & 	12.091753	 & 	0.8 $\pm$ 0.2	 & 	166 $\pm$ 7 \\ 
96	 & 	83.036903	 & 	12.111020	 & 	1.6 $\pm$ 0.4	 & 	166 $\pm$ 7 \\ 
97	 & 	83.037979	 & 	12.323105	 & 	0.6 $\pm$ 0.1	 & 	174 $\pm$ 8 \\ 
98	 & 	83.040497	 & 	12.442880	 & 	1.7 $\pm$ 0.2	 & 	16 $\pm$ 5 \\ 
99	 & 	83.040894	 & 	12.451411	 & 	0.6 $\pm$ 0.1	 & 	12 $\pm$ 7 \\ 
100	 & 	83.044777	 & 	12.450017	 & 	2.6 $\pm$ 0.4	 & 	40 $\pm$ 4 \\ 
\hline
\end{tabular}
\end{tabular}
\end{center}
\end{table*}
\begin{table*}
\begin{center}
\contcaption{Table A.2. continued.}
\label{table:BRC18}
\begin{tabular}{cc}
\begin{tabular}[t]{ccccr}  \hline
Star  & $\alpha$ (J2000)  & $\delta$ (J2000)  & P $\pm$ $\epsilon_P$ & $\theta$ $\pm$ $\epsilon_{\theta}$  \\ 
Id  &($\degree$)&($\degree$)& (\%) &($\degree$) \\\hline  
101	 & 	83.057884	 & 	12.316232	 & 	0.6 $\pm$ 0.2	 & 	128 $\pm$ 7 \\ 
102	 & 	83.057983	 & 	12.409883	 & 	2.3 $\pm$ 0.3	 & 	47 $\pm$ 4 \\ 
103	 & 	83.068832	 & 	12.094130	 & 	0.8 $\pm$ 0.3	 & 	171 $\pm$ 10 \\
104	 & 	83.073029	 & 	12.106179	 & 	0.8 $\pm$ 0.3	 & 	161 $\pm$ 10 \\
105	 & 	83.075089	 & 	12.124679	 & 	0.4 $\pm$ 0.2	 & 	157 $\pm$ 11 \\
\hline
\multicolumn{5}{c}{\bf BRC 18}\\
1	 & 	85.996071	 & 	9.090541	 & 	2.8 $\pm$ 0.2	 & 	172 $\pm$ 4 \\ 
2	 & 	86.008682	 & 	9.260374	 & 	2.2 $\pm$ 0.3	 & 	95 $\pm$ 4 \\ 
3	 & 	86.010780	 & 	9.102589	 & 	1.1 $\pm$ 0.4	 & 	174 $\pm$ 9 \\ 
4	 & 	86.017776	 & 	9.253828	 & 	1.8 $\pm$ 0.3	 & 	179 $\pm$ 5 \\ 
5	 & 	86.018143	 & 	9.157030	 & 	1.4 $\pm$ 0.3	 & 	1 $\pm$ 6 \\ 
6	 & 	86.032570	 & 	9.136529	 & 	1.8 $\pm$ 0.3	 & 	179 $\pm$ 5 \\ 
7	 & 	86.036148	 & 	9.275244	 & 	1.7 $\pm$ 0.2	 & 	3 $\pm$ 5 \\  
8	 & 	86.037628	 & 	9.082024	 & 	1.3 $\pm$ 0.4	 & 	8 $\pm$ 9 \\ 
9	 & 	86.040085	 & 	9.070679	 & 	1.8 $\pm$ 0.5	 & 	2 $\pm$ 9 \\ 
10	 & 	86.045395	 & 	9.290759	 & 	2.3 $\pm$ 0.4	 & 	4 $\pm$ 4 \\ 
11	 & 	86.045616	 & 	9.269732	 & 	2.3 $\pm$ 0.6	 & 	172 $\pm$ 7 \\ 
12	 & 	86.045921	 & 	9.138239	 & 	1.3 $\pm$ 0.2	 & 	166 $\pm$ 5 \\ 
13	 & 	86.046082	 & 	9.238827	 & 	2.2 $\pm$ 0.4	 & 	178 $\pm$ 5 \\ 
14	 & 	86.054008	 & 	9.077346	 & 	1.8 $\pm$ 0.3	 & 	168 $\pm$ 5 \\ 
15	 & 	86.056496	 & 	9.113747	 & 	1.9 $\pm$ 0.3	 & 	170 $\pm$ 5 \\ 
16	 & 	86.058495	 & 	9.155698	 & 	1.2 $\pm$ 0.2	 & 	175 $\pm$ 5 \\ 
17	 & 	86.058922	 & 	9.060800	 & 	1.1 $\pm$ 0.3	 & 	170 $\pm$ 9 \\ 
18	 & 	86.060120	 & 	9.321529	 & 	2.6 $\pm$ 0.4	 & 	174 $\pm$ 5 \\ 
19	 & 	86.063675	 & 	9.083546	 & 	2.2 $\pm$ 0.6	 & 	1 $\pm$ 8 \\ 
20	 & 	86.065529	 & 	9.365006	 & 	0.3 $\pm$ 0.1	 & 	176 $\pm$ 8 \\ 
21	 & 	86.069878	 & 	9.358155	 & 	1.9 $\pm$ 0.4	 & 	175 $\pm$ 5 \\ 
22	 & 	86.072441	 & 	9.318766	 & 	1.8 $\pm$ 0.4	 & 	175 $\pm$ 7 \\ 
23	 & 	86.074135	 & 	9.343622	 & 	2.0 $\pm$ 0.2	 & 	166 $\pm$ 4 \\ 
24	 & 	86.075256	 & 	9.294579	 & 	1.1 $\pm$ 0.5	 & 	173 $\pm$ 11 \\ 25	 & 	86.078064	 & 	9.356351	 & 	1.6 $\pm$ 0.3	 & 	166 $\pm$ 5 \\ 
26	 & 	86.080666	 & 	9.089832	 & 	0.9 $\pm$ 0.4	 & 	150 $\pm$ 11 \\
27	 & 	86.082092	 & 	9.220636	 & 	1.7 $\pm$ 0.1	 & 	179 $\pm$ 4 \\ 
28	 & 	86.082748	 & 	9.072447	 & 	1.8 $\pm$ 0.5	 & 	1 $\pm$ 7 \\ 
29	 & 	86.086159	 & 	9.306302	 & 	2.1 $\pm$ 0.3	 & 	162 $\pm$ 4 \\ 
30	 & 	86.087486	 & 	9.308671	 & 	1.7 $\pm$ 0.5	 & 	162 $\pm$ 7 \\ 
31	 & 	86.087769	 & 	9.349596	 & 	0.6 $\pm$ 0.2	 & 	169 $\pm$ 9 \\ 
32	 & 	86.089050	 & 	9.308807	 & 	4.8 $\pm$ 0.6	 & 	119 $\pm$ 3 \\ 
33	 & 	86.094055	 & 	9.265513	 & 	1.9 $\pm$ 0.3	 & 	172 $\pm$ 4 \\ 
34	 & 	86.094940	 & 	9.237477	 & 	1.8 $\pm$ 0.3	 & 	170 $\pm$ 5 \\ 
35	 & 	86.103493	 & 	9.391103	 & 	1.3 $\pm$ 0.2	 & 	167 $\pm$ 6 \\ 
36	 & 	86.103500	 & 	9.231409	 & 	2.2 $\pm$ 0.3	 & 	176 $\pm$ 5 \\ 
37	 & 	86.105469	 & 	9.091146	 & 	2.3 $\pm$ 0.4	 & 	178 $\pm$ 4 \\ 
38	 & 	86.112007	 & 	9.250659	 & 	1.7 $\pm$ 0.2	 & 	170 $\pm$ 5 \\ 
39	 & 	86.118973	 & 	9.206998	 & 	1.7 $\pm$ 0.3	 & 	171 $\pm$ 6 \\ 
40	 & 	86.120583	 & 	9.294956	 & 	1.7 $\pm$ 0.2	 & 	175 $\pm$ 5 \\ 
41	 & 	86.123268	 & 	9.218368	 & 	2.4 $\pm$ 0.3	 & 	169 $\pm$ 4 \\ 
42	 & 	86.124565	 & 	9.041305	 & 	1.2 $\pm$ 0.4	 & 	3 $\pm$ 12 \\ 
43	 & 	86.124680	 & 	9.213189	 & 	1.9 $\pm$ 0.5	 & 	163 $\pm$ 7 \\ 
44	 & 	86.127502	 & 	9.218301	 & 	3.5 $\pm$ 0.2	 & 	1 $\pm$ 3 \\ 
45	 & 	86.128441	 & 	9.314024	 & 	2.0 $\pm$ 0.3	 & 	170 $\pm$ 4 \\ 
46	 & 	86.128914	 & 	8.990820	 & 	2.4 $\pm$ 0.4	 & 	1 $\pm$ 5 \\ 
47	 & 	86.129539	 & 	9.006216	 & 	2.7 $\pm$ 0.3	 & 	175 $\pm$ 5 \\ 
48	 & 	86.130562	 & 	9.378459	 & 	1.8 $\pm$ 0.3	 & 	179 $\pm$ 5 \\ 
49	 & 	86.130806	 & 	9.386317	 & 	1.9 $\pm$ 0.5	 & 	168 $\pm$ 8 \\ 
50	 & 	86.132439	 & 	8.970585	 & 	1.8 $\pm$ 0.2	 & 	174 $\pm$ 4 \\ 
\hline
\end{tabular}
\begin{tabular}[t]{ccccr}  \hline
Star  & $\alpha$ (J2000)  & $\delta$ (J2000)  & P $\pm$ $\epsilon_P$ & $\theta$ $\pm$ $\epsilon_{\theta}$  \\ 
Id  &($\degree$)&($\degree$)& (\%) &($\degree$) \\\hline 
51	 & 	86.132782	 & 	9.398228	 & 	2.0 $\pm$ 0.2	 & 	176 $\pm$ 3 \\ 
52	 & 	86.134117	 & 	9.344035	 & 	1.9 $\pm$ 0.2	 & 	171 $\pm$ 5 \\ 
53	 & 	86.134506	 & 	9.382899	 & 	0.7 $\pm$ 0.1	 & 	180 $\pm$ 6 \\ 
54	 & 	86.136421	 & 	9.320812	 & 	2.0 $\pm$ 0.3	 & 	161 $\pm$ 4 \\ 
55	 & 	86.136726	 & 	9.380013	 & 	1.2 $\pm$ 0.1	 & 	169 $\pm$ 4 \\ 
56	 & 	86.137741	 & 	9.393141	 & 	1.8 $\pm$ 0.2	 & 	171 $\pm$ 4 \\ 
57	 & 	86.141449	 & 	9.300094	 & 	2.0 $\pm$ 0.3	 & 	172 $\pm$ 5 \\ 
58	 & 	86.147583	 & 	9.043317	 & 	1.9 $\pm$ 0.4	 & 	173 $\pm$ 6 \\ 
59	 & 	86.156700	 & 	8.953581	 & 	2.7 $\pm$ 0.4	 & 	168 $\pm$ 4 \\ 
60	 & 	86.157463	 & 	9.273091	 & 	2.7 $\pm$ 0.1	 & 	176 $\pm$ 4 \\ 
61	 & 	86.157883	 & 	9.270731	 & 	4.9 $\pm$ 2.2	 & 	156 $\pm$ 13 \\
62	 & 	86.163757	 & 	9.006329	 & 	1.2 $\pm$ 0.4	 & 	165 $\pm$ 9 \\ 
63	 & 	86.163773	 & 	9.327895	 & 	2.1 $\pm$ 0.5	 & 	159 $\pm$ 7 \\ 
64	 & 	86.166702	 & 	8.995458	 & 	1.8 $\pm$ 0.2	 & 	1 $\pm$ 5 \\ 
65	 & 	86.171822	 & 	9.409552	 & 	2.6 $\pm$ 0.6	 & 	173 $\pm$ 7 \\ 
66	 & 	86.173531	 & 	9.021292	 & 	2.0 $\pm$ 0.5	 & 	171 $\pm$ 8 \\ 
67	 & 	86.176399	 & 	9.426946	 & 	1.1 $\pm$ 0.2	 & 	13 $\pm$ 7 \\ 
68	 & 	86.179832	 & 	8.990156	 & 	2.0 $\pm$ 0.4	 & 	174 $\pm$ 6 \\ 
69	 & 	86.181885	 & 	9.288206	 & 	3.4 $\pm$ 0.2	 & 	15 $\pm$ 2 \\ 
70	 & 	86.183594	 & 	9.433202	 & 	2.3 $\pm$ 0.4	 & 	176 $\pm$ 5 \\ 
71	 & 	86.184059	 & 	9.054083	 & 	2.8 $\pm$ 0.5	 & 	176 $\pm$ 5 \\ 
72	 & 	86.184166	 & 	9.290176	 & 	2.1 $\pm$ 0.4	 & 	3 $\pm$ 5 \\ 
73	 & 	86.184479	 & 	9.020583	 & 	1.7 $\pm$ 0.7	 & 	169 $\pm$ 12 \\
74	 & 	86.184624	 & 	9.024846	 & 	1.8 $\pm$ 0.6	 & 	167 $\pm$ 10 \\
75	 & 	86.186218	 & 	9.217989	 & 	1.0 $\pm$ 0.3	 & 	74 $\pm$ 7 \\ 
76	 & 	86.189857	 & 	9.426837	 & 	2.0 $\pm$ 0.2	 & 	175 $\pm$ 3 \\ 
77	 & 	86.197639	 & 	8.984843	 & 	1.3 $\pm$ 0.4	 & 	153 $\pm$ 8 \\ 
78	 & 	86.198196	 & 	9.247959	 & 	2.8 $\pm$ 0.3	 & 	3 $\pm$ 3 \\ 
79	 & 	86.198570	 & 	9.345872	 & 	1.8 $\pm$ 0.2	 & 	165 $\pm$ 5 \\ 
80	 & 	86.198761	 & 	9.389240	 & 	2.3 $\pm$ 0.2	 & 	168 $\pm$ 4 \\ 
81	 & 	86.201401	 & 	9.402506	 & 	2.1 $\pm$ 0.4	 & 	3 $\pm$ 5 \\ 
82	 & 	86.201805	 & 	8.962544	 & 	1.5 $\pm$ 0.2	 & 	168 $\pm$ 5 \\ 
83	 & 	86.201859	 & 	8.962573	 & 	1.8 $\pm$ 0.3	 & 	175 $\pm$ 5 \\ 
84	 & 	86.204926	 & 	9.022941	 & 	0.9 $\pm$ 0.2	 & 	147 $\pm$ 6 \\ 
85	 & 	86.206192	 & 	9.020267	 & 	1.5 $\pm$ 0.2	 & 	164 $\pm$ 4 \\ 
86	 & 	86.206825	 & 	9.378925	 & 	1.6 $\pm$ 0.3	 & 	174 $\pm$ 5 \\ 
87	 & 	86.220009	 & 	9.338388	 & 	1.3 $\pm$ 0.2	 & 	12 $\pm$ 6 \\ 
88	 & 	86.223259	 & 	8.902828	 & 	1.7 $\pm$ 0.3	 & 	179 $\pm$ 6 \\ 
89	 & 	86.231506	 & 	9.266651	 & 	3.4 $\pm$ 0.7	 & 	171 $\pm$ 6 \\ 
90	 & 	86.233208	 & 	8.912957	 & 	1.8 $\pm$ 0.2	 & 	170 $\pm$ 4 \\ 
91	 & 	86.236847	 & 	9.377953	 & 	2.0 $\pm$ 0.3	 & 	165 $\pm$ 5 \\ 
92	 & 	86.237694	 & 	8.960582	 & 	0.6 $\pm$ 0.1	 & 	153 $\pm$ 6 \\ 
93	 & 	86.239525	 & 	9.395449	 & 	2.2 $\pm$ 0.2	 & 	168 $\pm$ 3 \\ 
94	 & 	86.241272	 & 	9.356718	 & 	0.7 $\pm$ 0.3	 & 	3 $\pm$ 12 \\ 
95	 & 	86.242500	 & 	9.450074	 & 	2.8 $\pm$ 0.3	 & 	173 $\pm$ 3 \\ 
96	 & 	86.242744	 & 	9.059106	 & 	4.5 $\pm$ 0.5	 & 	141 $\pm$ 3 \\ 
97	 & 	86.242882	 & 	9.232504	 & 	2.3 $\pm$ 0.5	 & 	176 $\pm$ 6 \\ 
98	 & 	86.243347	 & 	9.365779	 & 	1.8 $\pm$ 0.6	 & 	172 $\pm$ 10 \\
99	 & 	86.245682	 & 	8.935081	 & 	1.4 $\pm$ 0.3	 & 	174 $\pm$ 6 \\ 
100	 & 	86.246887	 & 	9.059897	 & 	3.1 $\pm$ 0.2	 & 	140 $\pm$ 4 \\ 
101	 & 	86.248924	 & 	8.915370	 & 	3.1 $\pm$ 0.5	 & 	173 $\pm$ 5 \\ 
102	 & 	86.249535	 & 	9.167620	 & 	1.1 $\pm$ 0.4	 & 	171 $\pm$ 12 \\
103	 & 	86.258911	 & 	9.155283	 & 	1.2 $\pm$ 0.2	 & 	172 $\pm$ 6 \\ 
104	 & 	86.260796	 & 	9.063945	 & 	2.5 $\pm$ 0.1	 & 	136 $\pm$ 6 \\ 
105	 & 	86.261047	 & 	9.430109	 & 	1.5 $\pm$ 0.3	 & 	170 $\pm$ 6 \\ 
106	 & 	86.265808	 & 	9.169325	 & 	1.3 $\pm$ 0.4	 & 	137 $\pm$ 7 \\ 
\hline
\end{tabular}
\end{tabular}
\end{center}
\end{table*}
\begin{table*}
\begin{center}
\contcaption{Table A.2. continued.}
\label{table:BRC18}
\begin{tabular}{cc}
\begin{tabular}[t]{ccccr}  \hline
Star  & $\alpha$ (J2000)  & $\delta$ (J2000)  & P $\pm$ $\epsilon_P$ & $\theta$ $\pm$ $\epsilon_{\theta}$  \\ 
Id  &($\degree$)&($\degree$)& (\%) &($\degree$) \\\hline  
107	 & 	86.270256	 & 	9.068852	 & 	2.0 $\pm$ 0.4	 & 	135 $\pm$ 5 \\ 
108	 & 	86.270340	 & 	9.258509	 & 	1.5 $\pm$ 0.3	 & 	2 $\pm$ 6 \\ 
109	 & 	86.273323	 & 	9.187541	 & 	1.5 $\pm$ 0.6	 & 	179 $\pm$ 12 \\
110	 & 	86.275368	 & 	9.218266	 & 	1.1 $\pm$ 0.3	 & 	173 $\pm$ 8 \\ 
111	 & 	86.277069	 & 	9.167779	 & 	1.1 $\pm$ 0.4	 & 	170 $\pm$ 11 \\
112	 & 	86.278572	 & 	9.396746	 & 	1.7 $\pm$ 0.2	 & 	171 $\pm$ 5 \\ 
113	 & 	86.286713	 & 	9.218583	 & 	1.6 $\pm$ 0.2	 & 	7 $\pm$ 5 \\ 
114	 & 	86.287544	 & 	8.930645	 & 	1.3 $\pm$ 0.2	 & 	166 $\pm$ 4 \\ 
115	 & 	86.293060	 & 	9.198799	 & 	1.7 $\pm$ 0.2	 & 	179 $\pm$ 5 \\ 
116	 & 	86.295547	 & 	8.942298	 & 	1.0 $\pm$ 0.3	 & 	173 $\pm$ 9 \\ 
117	 & 	86.296295	 & 	8.915566	 & 	1.3 $\pm$ 0.2	 & 	175 $\pm$ 5 \\ 
118	 & 	86.301781	 & 	8.936301	 & 	1.3 $\pm$ 0.3	 & 	173 $\pm$ 7 \\ 
119	 & 	86.302681	 & 	8.998691	 & 	1.9 $\pm$ 0.3	 & 	143 $\pm$ 4 \\ 
120	 & 	86.307884	 & 	9.078981	 & 	1.5 $\pm$ 0.1	 & 	124 $\pm$ 5 \\ 
121	 & 	86.310135	 & 	8.929467	 & 	1.2 $\pm$ 0.3	 & 	166 $\pm$ 6 \\ 
122	 & 	86.312378	 & 	9.302844	 & 	1.9 $\pm$ 0.4	 & 	166 $\pm$ 5 \\ 
123	 & 	86.312469	 & 	9.454061	 & 	2.1 $\pm$ 0.3	 & 	176 $\pm$ 4 \\ 
124	 & 	86.313522	 & 	9.272567	 & 	1.4 $\pm$ 0.1	 & 	4 $\pm$ 6 \\ 
125	 & 	86.323395	 & 	9.453463	 & 	1.3 $\pm$ 0.4	 & 	6 $\pm$ 9 \\ 
126	 & 	86.328117	 & 	9.445957	 & 	1.1 $\pm$ 0.2	 & 	177 $\pm$ 7 \\ 
127	 & 	86.329376	 & 	9.415783	 & 	1.4 $\pm$ 0.2	 & 	174 $\pm$ 5 \\ 
128	 & 	86.332329	 & 	9.228611	 & 	0.9 $\pm$ 0.1	 & 	177 $\pm$ 6 \\ 
129	 & 	86.334602	 & 	9.283645	 & 	1.0 $\pm$ 0.2	 & 	172 $\pm$ 6 \\ 
130	 & 	86.334671	 & 	9.083313	 & 	1.6 $\pm$ 0.4	 & 	173 $\pm$ 7 \\ 
131	 & 	86.335464	 & 	9.038640	 & 	2.3 $\pm$ 0.4	 & 	133 $\pm$ 5 \\ 
132	 & 	86.336929	 & 	9.465528	 & 	1.9 $\pm$ 0.3	 & 	176 $\pm$ 4 \\ 
133	 & 	86.338112	 & 	9.311886	 & 	1.2 $\pm$ 0.1	 & 	173 $\pm$ 5 \\ 
134	 & 	86.343285	 & 	9.242063	 & 	0.5 $\pm$ 0.1	 & 	38 $\pm$ 8 \\ 
135	 & 	86.346771	 & 	9.030640	 & 	1.3 $\pm$ 0.4	 & 	165 $\pm$ 9 \\ 
136	 & 	86.351921	 & 	9.396746	 & 	1.4 $\pm$ 0.3	 & 	169 $\pm$ 7 \\ 
137	 & 	86.353699	 & 	9.426005	 & 	1.3 $\pm$ 0.2	 & 	165 $\pm$ 6 \\ 
138	 & 	86.363251	 & 	9.257471	 & 	2.8 $\pm$ 0.9	 & 	16 $\pm$ 10 \\ 
139	 & 	86.363678	 & 	9.417241	 & 	1.4 $\pm$ 0.6	 & 	170 $\pm$ 11 \\
140	 & 	86.371529	 & 	9.181946	 & 	1.1 $\pm$ 0.4	 & 	158 $\pm$ 9 \\ 
141	 & 	86.373787	 & 	8.945542	 & 	1.1 $\pm$ 0.5	 & 	150 $\pm$ 10 \\
142	 & 	86.374512	 & 	9.408002	 & 	1.0 $\pm$ 0.3	 & 	159 $\pm$ 7 \\ 
143	 & 	86.375862	 & 	8.947497	 & 	1.2 $\pm$ 0.5	 & 	166 $\pm$ 11 \\
144	 & 	86.378395	 & 	9.299932	 & 	0.9 $\pm$ 0.1	 & 	169 $\pm$ 6 \\ 
145	 & 	86.378502	 & 	9.133583	 & 	2.8 $\pm$ 0.5	 & 	168 $\pm$ 5 \\ 
146	 & 	86.386871	 & 	9.399343	 & 	1.5 $\pm$ 0.3	 & 	176 $\pm$ 5 \\ 
147	 & 	86.388138	 & 	9.417358	 & 	2.3 $\pm$ 0.4	 & 	160 $\pm$ 4 \\ 
148	 & 	86.390251	 & 	8.931575	 & 	0.8 $\pm$ 0.3	 & 	176 $\pm$ 10 \\
149	 & 	86.392273	 & 	9.261717	 & 	0.9 $\pm$ 0.1	 & 	17 $\pm$ 6 \\ 
150	 & 	86.393578	 & 	9.217236	 & 	1.3 $\pm$ 0.2	 & 	0 $\pm$ 5 \\ 
151	 & 	86.394753	 & 	9.177710	 & 	1.2 $\pm$ 0.3	 & 	14 $\pm$ 8 \\ 
152	 & 	86.398598	 & 	9.418167	 & 	1.9 $\pm$ 0.4	 & 	164 $\pm$ 5 \\ 
153	 & 	86.409279	 & 	8.944745	 & 	0.7 $\pm$ 0.2	 & 	158 $\pm$ 7 \\ 
154	 & 	86.411713	 & 	9.411193	 & 	2.1 $\pm$ 0.4	 & 	166 $\pm$ 6 \\ 
155	 & 	86.414032	 & 	9.420681	 & 	2.0 $\pm$ 0.2	 & 	158 $\pm$ 4 \\ 
\hline
\end{tabular}
\begin{tabular}[t]{ccccr}  \hline
Star  & $\alpha$ (J2000)  & $\delta$ (J2000)  & P $\pm$ $\epsilon_P$ & $\theta$ $\pm$ $\epsilon_{\theta}$  \\ 
Id  &($\degree$)&($\degree$)& (\%) &($\degree$) \\\hline 
156	 & 	86.415398	 & 	8.945398	 & 	1.4 $\pm$ 0.3	 & 	177 $\pm$ 5 \\ 
157	 & 	86.422096	 & 	9.155535	 & 	1.1 $\pm$ 0.3	 & 	123 $\pm$ 5 \\ 
158	 & 	86.425629	 & 	9.407340	 & 	1.1 $\pm$ 0.1	 & 	177 $\pm$ 5 \\ 
159	 & 	86.425873	 & 	9.157034	 & 	1.9 $\pm$ 0.3	 & 	169 $\pm$ 5 \\ 
160	 & 	86.426407	 & 	9.129604	 & 	0.8 $\pm$ 0.1	 & 	169 $\pm$ 5 \\ 
161	 & 	86.427895	 & 	9.289151	 & 	3.2 $\pm$ 0.9	 & 	118 $\pm$ 8 \\ 
162	 & 	86.429314	 & 	9.167664	 & 	1.0 $\pm$ 0.3	 & 	166 $\pm$ 9 \\ 
163	 & 	86.435654	 & 	9.163746	 & 	0.7 $\pm$ 0.2	 & 	175 $\pm$ 9 \\ 
164	 & 	86.444221	 & 	9.210876	 & 	1.4 $\pm$ 0.1	 & 	173 $\pm$ 6 \\ 
165	 & 	86.445435	 & 	9.206098	 & 	1.1 $\pm$ 0.1	 & 	178 $\pm$ 5 \\ 
166	 & 	86.447556	 & 	9.193343	 & 	0.7 $\pm$ 0.3	 & 	166 $\pm$ 9 \\ 
167	 & 	86.448502	 & 	9.191902	 & 	1.2 $\pm$ 0.2	 & 	175 $\pm$ 7 \\ 
168	 & 	86.450058	 & 	8.928351	 & 	1.1 $\pm$ 0.2	 & 	175 $\pm$ 5 \\ 
169	 & 	86.450966	 & 	8.983732	 & 	1.7 $\pm$ 0.3	 & 	10 $\pm$ 6 \\ 
170	 & 	86.451347	 & 	9.174396	 & 	1.2 $\pm$ 0.2	 & 	3 $\pm$ 7 \\ 
171	 & 	86.458473	 & 	9.167095	 & 	1.4 $\pm$ 0.3	 & 	179 $\pm$ 6 \\ \hline
\end{tabular}
\end{tabular}
\end{center}  
\end{table*}

\end{appendix}


\end{document}